\documentclass[aps,prx,twocolumn,superscriptaddress,amsmath,amssymb,floatfix]{revtex4-2}

\usepackage{graphicx}
\usepackage{dcolumn}
\usepackage{bm}
\usepackage{xcolor}
\usepackage{hyperref}
\usepackage{braket}
\usepackage{siunitx}
\usepackage{physics}

\hypersetup{
    colorlinks=true,
    linkcolor=black,
    citecolor=black,
    urlcolor=black
}

\newif\ifshowrevisions
\showrevisionsfalse

\DeclareRobustCommand{\revb}[1]{\ifshowrevisions{\color{blue}#1}\else#1\fi}
\definecolor{revpurple}{rgb}{0.50,0.00,0.55}

\newenvironment{revcblock}{\ifshowrevisions\color{revpurple}\fi}{}
\DeclareRobustCommand{\revd}[1]{#1}
\DeclareRobustCommand{\revpzero}[1]{#1}

\DeclareRobustCommand{\reve}[1]{#1}
\newenvironment{reveblock}{}{}
\DeclareRobustCommand{\revnow}[1]{#1}
\newenvironment{revnowblock}{}{}
\DeclareRobustCommand{\revsym}[1]{#1}

\begin{document}

\title{\texorpdfstring{\revnow{An exchange-assisted entangling gate between $^{87}\mathrm{Rb}$ and $^{171}\mathrm{Yb}$ Rydberg atoms}}{An exchange-assisted entangling gate between 87Rb and 171Yb Rydberg atoms}}

\author{Han Wang}
\affiliation{QudeLeap Research, Shanghai 200030, China}
\affiliation{The Hong Kong University of Science and Technology (Guangzhou), Guangdong 511453, China}

\author{Erdong Huang}
\affiliation{QudeLeap Research, Shanghai 200030, China}
\affiliation{The Hong Kong University of Science and Technology (Guangzhou), Guangdong 511453, China}

\author{Mingrui Jing}
\affiliation{QudeLeap Research, Shanghai 200030, China}
\affiliation{The Hong Kong University of Science and Technology (Guangzhou), Guangdong 511453, China}

\author{Hongshun Yao}
\affiliation{QudeLeap Research, Shanghai 200030, China}
\affiliation{The Hong Kong University of Science and Technology (Guangzhou), Guangdong 511453, China}

\author{Xin Wang}
\email{wangxinfelix@gmail.com}
\affiliation{The Hong Kong University of Science and Technology (Guangzhou), Guangdong 511453, China}

\author{Jin-Guo Liu}
\email{jinguoliu@hkust-gz.edu.cn}
\affiliation{The Hong Kong University of Science and Technology (Guangzhou), Guangdong 511453, China}

\date{\today}

\begin{abstract}
\revnow{Neutral-atom tweezer arrays support scalable quantum information processing. Dual-species $^{87}\mathrm{Rb}$--$^{171}\mathrm{Yb}$ arrays combine long-lived ytterbium nuclear-spin data qubits with fast, species-selective rubidium ancilla control and readout.
However, realizing interspecies gates without inducing destructive Stark mixing in divalent atoms remains an outstanding problem. Here, we identify an optically accessible $S{+}S\leftrightarrow P{+}P$ F\"orster resonance at zero electric field, providing strong dipole-dipole exchange at array pitch. Using a shaped optical pulse under finite control response, we demonstrate a $0.36\,\mu\mathrm{s}$ exchange-assisted controlled-$Z$ gate with an intrinsic fidelity of $99.91\%$, remaining above $99.85\%$ under bounded perturbations. We also identify an auxiliary repulsive van der Waals channel, providing a comprehensive toolbox for hybrid quantum processors.}
\end{abstract}

\maketitle

\section{Introduction}
\label{sec:intro}

\begin{figure*}[t]
\centering
\includegraphics[width=\textwidth]{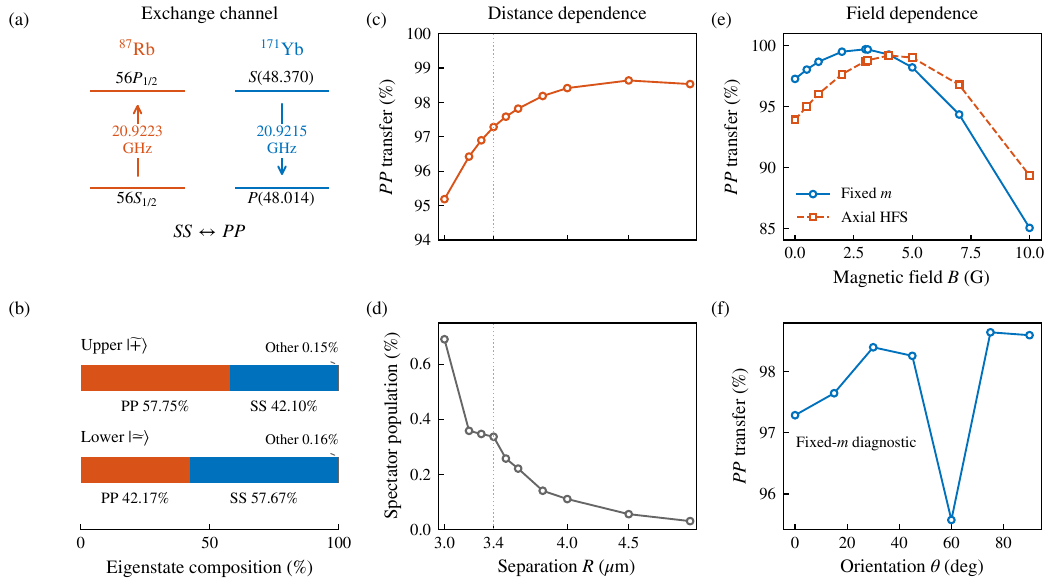}
\caption{\textbf{Microscopic characterization of the $^{87}\mathrm{Rb}$--$^{171}\mathrm{Yb}$ F\"orster exchange channel.}
(a)~Species-resolved energy-level diagram for the resonant exchange interaction between $\mathrm{Rb}\,(56S_{1/2}\to 56P_{1/2})$ and $\mathrm{Yb}\,[S(\nu\approx 48.37)\to P(\nu\approx 48.01)]$, yielding sub-megahertz unperturbed F\"orster defects $\Delta_{\mathrm{el}}$ and $\Delta_{\mathrm{hfs}}$ in the zero-field limit ($B=0, E=0$).
The level positions are schematic; the labeled transition energies identify the near-resonant exchange.
(b)~Stacked compositions of the two bright eigenstates $\lvert\widetilde{\pm}\rangle$ at $R=3.4\,\mu\mathrm{m}$ ($B=0$).
Each bar sums to $100\%$: orange denotes $PP$, blue denotes $SS$, and gray denotes spectator states. Component percentages are labeled directly.
Both eigenstates have more than $99.8\%$ weight within the target $\{\lvert PP\rangle, \lvert SS\rangle\}$ subspace, with off-resonant spectator states contributing only $0.15$--$0.16\%$.
(c,d)~First-exchange transfer into $\lvert PP\rangle$ and simultaneous spectator population versus separation $R$, at zero magnetic field ($B=0, E=0$) and $\theta=0$. The dotted line marks $R=3.4\,\mu\mathrm{m}$.
(e)~First-exchange transfer versus magnetic field: blue circles denote the fixed electronic-$m$ reference and orange squares the axial hyperfine-resolved basis.
(f)~Orientation dependence in the fixed-$m$ reference model.
The field scan compares the fixed-$m$ and axial hyperfine-resolved models; the separate angle scan diagnoses anisotropy only within the fixed-$m$ model. Markers denote calculated points, with lines as guides to the eye.
The entangling gate protocol in Sec.~\ref{sec:forster_gate} operates at $\theta=0$ and $B=3.10\,\mathrm{G}$.}
\label{fig:forster_channel}
\end{figure*}

\begin{revnowblock}
Reconfigurable optical tweezer arrays support quantum simulation, entangling gates, and logical operations with individually controlled neutral atoms~\cite{Browaeys2020,Kaufman2021,Evered2023,Bluvstein2024,Evered2026}.
Repeated quantum error correction also requires selective ancilla measurement and reset while preserving the data register.
In a single-species array, readout photons can perturb nearby data atoms that share the same optical transitions.
Assigning the data and ancilla roles to different species provides spectral separation for selective control and measurement~\cite{Beterov2015,Singh2022,Sheng2022,Anand2024,Wang2026}.
Two-qubit entangling gates also require strong, switchable interactions between the species.
In dual-alkali architectures, notably $\mathrm{Rb}$--$\mathrm{Cs}$, interspecies gates have been realized by exploiting electric-field-tuned F\"orster resonances between Rydberg states~\cite{Anand2024,Miles2026,Ireland2024}.
Extending this strategy to an alkali--divalent mixture requires optically accessible pair channels with limited spectator mixing in the dense Rydberg spectrum of the two-electron atom.

Pairing an alkali atom ($^{87}\mathrm{Rb}$) with an alkaline-earth-like atom ($^{171}\mathrm{Yb}$) offers complementary control and storage capabilities for fault-tolerant quantum computing~\cite{Wu2022,Ma2023,Peper2024,Zhang2025Dual}.
The $I=1/2$ nuclear spin of divalent $^{171}\mathrm{Yb}$ provides a weakly magnetically sensitive data-qubit manifold in the ground $^{1}S_0$ and metastable clock $^{3}P_0$ states. Dominant Rydberg decay channels populate states outside the computational subspace, allowing optical detection and erasure conversion before causing Pauli errors~\cite{Wu2022,Ma2023,Peper2024}.
$^{87}\mathrm{Rb}$ can act as the ancilla, with established near-infrared control, rapid transport, and fast readout at $780\,\mathrm{nm}$, spectrally isolated from ytterbium's optical transitions~\cite{Wu2022,Weber2026,DAMOP2026_Bao,DAMOP2025_Zheng}.
Zhang, Arunseangroj, and Xu recently proposed a hybrid architecture combining Yb data atoms with Rb ensembles for local control and rapid stabilizer measurements~\cite{Zhang2025Dual}.
Their model uses generic resonant pair states with an assumed interaction strength, leaving the microscopic channel unspecified.
Experimental programs have targeted Rb--Yb pair spectroscopy~\cite{DAMOP2025_Zheng}, and a recent conference abstract reports interspecies F\"orster-resonance characterization~\cite{DAMOP2026_Xu}.
These developments motivate a gate protocol based on a specified microscopic channel.

A useful gate channel must satisfy several requirements simultaneously.
The addressed states must be optically accessible, their exchange coupling must be strong enough for dynamics faster than radiative decay, and spectator-state mixing must remain limited.
For Yb, coupled electronic series and hyperfine structure complicate both the level energies and the dipole matrix elements; validated multichannel quantum defect theory (MQDT) models now permit quantitative pair-interaction calculations~\cite{Peper2024,Kuroda2025,Mogerle2026}.
An exchange-assisted gate must drive the interacting and noninteracting computational branches along closed trajectories, returning their populations with a conditional $\pi$ phase. This requires pulse design that accounts for the finite control response, beyond identifying a near-degenerate pair.
Resonant exchange already underlies two-atom dark-state gates~\cite{Petrosyan2017}, microwave-coupled dipolar gates~\cite{Giudici2025}, and asymmetric driven-return schemes outside the strong-blockade regime~\cite{Cole2025}.
Here we connect this dynamical control to a specified Rb--Yb multichannel interaction.

In this work, we identify an optically accessible $S{+}S\leftrightarrow P{+}P$ F\"orster resonance between $\mathrm{Rb}(56S_{1/2})$ and $\mathrm{Yb}[S(\nu\approx 48.37)]$ that is naturally near-degenerate at zero electric field (Fig.~\ref{fig:forster_channel}), completely circumventing the severe Stark mixing that plagues divalent atoms under external bias fields.
Operating in an isolated stretched Zeeman manifold, a shaped optical pulse driving the ytterbium qubit steers the exchange-hybridized pair through an asymmetric revival trajectory, restoring atomic populations while imparting a conditional $\pi$ phase to execute a controlled-$Z$ ($\mathrm{CZ}$) gate (Figs.~\ref{fig:forster_excitation} and \ref{fig:forster_gate}).
At an array spacing of $3.4\,\mu\mathrm{m}$, the full multichannel model predicts an intrinsic gate fidelity of $99.91\%$ within $0.36\,\mu\mathrm{s}$ under finite optical control bandwidth, with a sampled minimum of $99.85\%$ across bounded position and laser-amplitude perturbations (Fig.~\ref{fig:forster_robustness}).

Section~\ref{sec:forster_channel} distinguishes the fixed-electronic-$m$ channel characterization from the hyperfine-resolved model used for the gate.
Section~\ref{sec:forster_gate} develops the optical protocol, fidelity metric, and finite-response waveform, then tests that waveform under specified perturbations.
The appendices provide the state dictionary and interaction model (Appendix~\ref{app:hyperfine_convergence}), numerical checks and the scope of the spectroscopy and motion studies (Appendix~\ref{app:gate_convergence}), and an auxiliary static van der Waals candidate (Appendix~\ref{sec:vdw}).

\end{revnowblock}

\section{Microscopic exchange channel}
\label{sec:forster_channel}

We first characterize the channel in a fixed-electronic-$m$ model without Rb nuclear spin, then compare it with the axial hyperfine-resolved model used for the gate.
The interspecies exchange interaction is mediated by the electric-dipole-coupled $S{+}S\leftrightarrow P{+}P$ F\"orster channel between $^{87}\mathrm{Rb}$ and $^{171}\mathrm{Yb}$, defined by the pair states
\begin{equation}
\begin{aligned}
\lvert SS\rangle &\equiv
\lvert 56S_{1/2}\rangle_{\mathrm{Rb}}
\otimes\lvert S(\nu_S)\rangle_{\mathrm{Yb}},\\[3pt]
\lvert PP\rangle &\equiv
\lvert 56P_{1/2}\rangle_{\mathrm{Rb}}
\otimes\lvert P(\nu_P)\rangle_{\mathrm{Yb}}.
\end{aligned}
\label{eq:sspp}
\end{equation}
Here $\nu_S\approx 48.37$ and $\nu_P\approx 48.01$ are the effective principal quantum numbers identifying the target Yb levels ($F=1/2$, with $L=0$ for $S$ and $L=1$ for $P$; exact MQDT roots indexed in Table~\ref{tab:state_dictionary}).
The database defines $\nu=\sqrt{\mathcal R/(E_{\mathrm{ion}}-E)}$, where $\mathcal R$ is the Rydberg energy and $E_{\mathrm{ion}}$ is the $^{171}\mathrm{Yb}^{+}$ ground-state threshold with core angular momentum $F_{\mathrm{core}}=1$.
We denote the interatomic separation by $R$ and the angle between the internuclear vector and the magnetic quantization axis by $\theta$.
All microscopic interactions and atomic matrix elements are calculated using PairInteraction 2.5.0~\cite{Mogerle2026,Weber2017} with the $\mathrm{Rb}$ v1.2~\cite{Li2003,Mack2011} and $\mathrm{Yb171\_mqdt}$ v1.4~\cite{Peper2024,Kuroda2025} databases (specific state selectors and conventions are documented in Appendix~\ref{app:hyperfine_convergence}).
The two atoms undergo mutually compensating transitions near $20.9\,\mathrm{GHz}$ [Fig.~\ref{fig:forster_channel}(a)].
Defining the F\"orster defect as \revsym{$\Delta_{\mathrm{F}}$}$=E_{SS}-E_{PP}$, the unperturbed electronic centroid defect without Rb nuclear spin is $\Delta_{\mathrm{el}}/h=-0.76\,\mathrm{MHz}$.
Including the stretched hyperfine components of $^{87}\mathrm{Rb}$ shifts the unperturbed defect at zero magnetic field ($B=0, E=0$) to
\begin{equation}
\begin{aligned}
\frac{\Delta_{\mathrm{hfs}}}{h}
&=\frac{\Delta_{\mathrm{el}}}{h}
+\frac{3}{4}\frac{A_{56S}-A_{56P_{1/2}}}{h}\\
&=-0.69\,\mathrm{MHz}.
\end{aligned}
\label{eq:defect}
\end{equation}
Both defect values are negative, placing $\lvert SS\rangle$ slightly below $\lvert PP\rangle$ in energy; the $\sim70\,\mathrm{kHz}$ difference reflects the $^{87}\mathrm{Rb}$ hyperfine contribution (Appendix~\ref{app:hyperfine_convergence}).
Within the candidate manifolds examined in Appendix~\ref{app:hyperfine_convergence}, this channel combines sub-megahertz asymptotic defects with two-level dominance, as quantified below.
Other near-degenerate candidates exhibited strong coupling to individual perturber states, illustrating the need to test the pair eigenstates as well as their asymptotic energies.

At an interatomic separation of $R=3.4\,\mu\mathrm{m}$, the electric dipole--dipole interaction couples $\lvert SS\rangle$ and $\lvert PP\rangle$ with a matrix element $\lvert V\rvert/h=15.4\,\mathrm{MHz}$, driving coherent, nonradiative population exchange between the two species.
This coupling exceeds either sub-megahertz defect, placing the pair in the strong-exchange regime without deliberate dc Stark tuning~\cite{Beterov2016,Ireland2024} and circumventing the Stark-induced multichannel mixing that reduces the target-state fraction in divalent ytterbium~\cite{Hummel2024,Peper2024}.
Gate sensitivity to residual uncompensated stray fields is evaluated in Appendix~\ref{app:bias_fields}.
For comparison, in the idealized two-state limit $\{\lvert PP\rangle,\lvert SS\rangle\}$ with detuning $\delta=\Delta_{\mathrm{el}}$, the strong coupling drives rapid exchange oscillations at a generalized frequency $\Delta\nu_2=30.8\,\mathrm{MHz}$.
Starting from $\lvert SS\rangle$, the exchange reaches a peak transfer efficiency of $P_{\max}^{(2)}=99.9\%$ into $\lvert PP\rangle$ at the half-period $t_{\max}^{(2)}\approx 16.2\,\mathrm{ns}$ (analytical expressions are given in Appendix~\ref{app:hyperfine_convergence}).

\revpzero{We next include the surrounding atomic spectrum.
Full numerical diagonalization of the 2411-state fixed-$m$ pair Hamiltonian at $R=3.4\,\mu\mathrm{m}$ [Fig.~\ref{fig:forster_channel}(b)] yields a bright-state splitting of $31.1\,\mathrm{MHz}$.
Both bright eigenstates retain more than $99.8\%$ probability weight within the target $\{\lvert SS\rangle, \lvert PP\rangle\}$ subspace (with $\approx 58\%/42\%$ hybridization and only $\sim 0.15\%$ spectator admixture), confirming that off-resonant background levels introduce only minor state mixing.}

\revpzero{In this fixed-$m$ multichannel basis at $R=3.4\,\mu\mathrm{m}$ ($B=0$), time-dependent propagation starting from $\lvert SS\rangle$ reaches its first exchange maximum at $16.0\,\mathrm{ns}$ [Fig.~\ref{fig:forster_channel}(c,d)], transferring $97.3\%$ into $\lvert PP\rangle$, leaving $2.4\%$ in $\lvert SS\rangle$, and leaking only $0.3\%$ into spectator states.
This transfer saturates the theoretical upper bound of $97.3\%$ set by the static eigenstate projection (Appendix~\ref{app:hyperfine_convergence}).
The separation scan shows a tradeoff between interaction strength and subspace isolation: over $R=3.4$--$4.5\,\mu\mathrm{m}$, the peak transfer increases from $97.3\%$ to $98.6\%$, while the bright-state splitting drops from $31.1\,\mathrm{MHz}$ to $13.4\,\mathrm{MHz}$.}
\revb{The selected spacing is comparable to the $3.3$--$4.5\,\mu\mathrm{m}$ separations used in Yb two-atom experiments~\cite{Peper2024}.
Co-aligned Rb--Yb tweezer arrays have also been demonstrated with $^{174}\mathrm{Yb}$~\cite{Weber2026}.
Systematic single-parameter basis expansions varying principal quantum number, orbital angular momentum, and pair-energy cutoffs shift the fixed-$m$ bright splitting by less than $10^{-3}\,\mathrm{MHz}$; the corresponding gate-model checks are detailed in Appendix~\ref{app:gate_convergence}.}

The applied magnetic field affects unperturbed state exchange and coherent gate dynamics differently.
\revpzero{We include all electronic magnetic sublevels and the $^{87}\mathrm{Rb}$ nuclear spin ($I=3/2$) in the conserved axial sector to account for magnetic degeneracy and hyperfine splittings. At zero magnetic field ($B=0$), the first-exchange transfer is $93.92\%$.
Applying a bias magnetic field lifts magnetic degeneracy and isolates the stretched components, increasing the transfer to $98.81\%$ at $3.10\,\mathrm{G}$ and a sampled peak of $99.20\%$ at $4\,\mathrm{G}$ [Fig.~\ref{fig:forster_channel}(e)].
Although differential Zeeman shifts alter the unperturbed pair detuning at this operating point ($\Delta_{\mathrm{F}}/h = +4.66\,\mathrm{MHz}$; Table~\ref{tab:state_dictionary}), the dipole--dipole coupling $\lvert V\rvert/h=15.4\,\mathrm{MHz}$ remains dominant ($2\lvert V\rvert \gg \Delta_{\mathrm{F}}$), preserving the strong-exchange regime while isolating the stretched target subspace.}
Because the magnetic field also tunes the dynamical phase acquired during the driven return, maximizing the static swap efficiency does not automatically maximize the \revd{modeled loss-aware gate fidelity}.
Section~\ref{sec:forster_gate} selects $R=3.4\,\mu\mathrm{m}$, $\theta=0$, and $B=3.10\,\mathrm{G}$ as the reference operating point for pulse construction, optimization, and error validation.
\revpzero{Details of the finite-field multichannel basis, the resulting bright eigenstate weights, and tensor dipole--dipole angular sensitivity are provided in Appendix~\ref{app:hyperfine_convergence}.}

\begin{revcblock}
\section{\texorpdfstring{Exchange-assisted controlled-$Z$ gate}{Exchange-assisted controlled-Z gate}}
\label{sec:forster_gate}

Section~\ref{sec:forster_channel} established a \revpzero{two-level-dominated $\lvert SS\rangle\leftrightarrow\lvert PP\rangle$ exchange channel} at $R=3.4\,\mu\mathrm{m}$.
The exchange-coupled pair states support an interspecies controlled-$Z$ ($\mathrm{CZ}$) gate using a shaped Yb pulse that optically accesses the $\lvert SS\rangle$ pair state.

\subsection{\revnow{Optical addressing and conditional revival}}

\reve{Each atomic species is selectively addressed via electric-dipole transitions [Fig.~\ref{fig:forster_excitation}].}
For $^{171}\mathrm{Yb}$, encoding the nuclear-spin qubit in the metastable odd-parity $^{3}P_0$ manifold enables direct single-photon $E1$ excitation to the even-parity $S$ Rydberg state~\cite{Peper2024,Muniz2024}.
A $\sigma^+$-polarized laser pulse at $302\,\mathrm{nm}$ couples the qubit state $\lvert 1\rangle_{\mathrm{Yb}}=\lvert{}^{3}P_0,F{=}1/2,m_F{=}-1/2\rangle$ to the target Rydberg level $\lvert S(\nu\approx 48.37),F{=}1/2,m_F{=}+1/2\rangle$, while leaving the orthogonal state $\lvert 0\rangle_{\mathrm{Yb}}=\lvert{}^{3}P_0,F{=}1/2,m_F{=}+1/2\rangle$ dark \reve{within this addressed manifold}.
\revnow{The calculated dipole $|d_{+1}|=2.39\times10^{-3}\,ea_0$ (Table~\ref{tab:state_dictionary}) gives a peak-command power requirement of $44.5\,\mathrm{mW}$ at $302\,\mathrm{nm}$ for $\Omega_C/2\pi=11.78\,\mathrm{MHz}$ and a Gaussian $1/e^2$ intensity radius of $12\,\mu\mathrm{m}$, before optical losses.
In single-atom calculations including neighboring Rydberg levels within $\pm100\,\mathrm{GHz}$, off-resonant spectator excitation stays below $1.3\times10^{-5}$ for the optimized target waveform. Residual polarization imperfections and technical optical fluctuations contribute to apparatus-dependent error budgets.}

For $^{87}\mathrm{Rb}$, we model excitation from $\lvert1\rangle_{\mathrm{Rb}}=\lvert5S_{1/2},F=2,m_F=+2\rangle$ to $\lvert56S_{1/2},m_J=+1/2,m_I=+3/2\rangle$ by an effective two-photon Rabi frequency of $5\,\mathrm{MHz}$.
The intended ladder uses $780\,\mathrm{nm}$ and $480\,\mathrm{nm}$ light through a detuned $5P_{3/2}$ intermediate state~\cite{Saffman2010,Graham2019,Levine2019}.
The orthogonal logical state is encoded in the clock sublevel $\lvert0\rangle_{\mathrm{Rb}} = \lvert5S_{1/2}, F=1, m_F=0\rangle$, which remains dark during the excitation sequence because of the $6.835\,\mathrm{GHz}$ ground-hyperfine splitting and optical selection rules.
The distinct excitation wavelengths permit species-selective addressing; the intended optical transitions do not directly drive the exchange partner $PP$.

\begin{figure}[t]
\centering
\includegraphics[width=0.9\columnwidth]{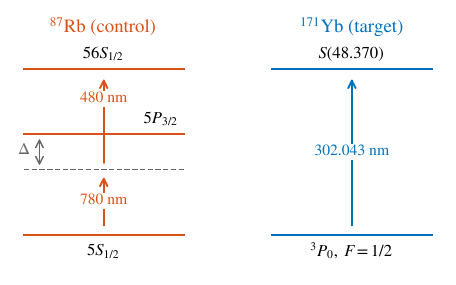}
\caption{\textbf{Optical excitation schemes for the $^{87}\mathrm{Rb}$--$^{171}\mathrm{Yb}$ gate.}
Level spacings and arrow lengths are schematic and not to scale.
Left (Rb control): two-photon ladder ($780\,\mathrm{nm}+480\,\mathrm{nm}$) excites $5S_{1/2}$ to $56S_{1/2}$ via the detuned $5P_{3/2}$ intermediate level (solid arrows: resonant fields; dashed: intermediate detuning).
Right (Yb target): single-photon $\sigma^+$ excitation at $302\,\mathrm{nm}$ couples the $^{3}P_0$ ($F{=}1/2$) clock state to the $S(\nu\approx 48.37)$ Rydberg state, leaving the partner state $\lvert PP\rangle$ unaddressed.}
\label{fig:forster_excitation}
\end{figure}

\begin{revnowblock}
In the computational basis $\lvert q_{\mathrm{Rb}}q_{\mathrm{Yb}}\rangle$, the gate has three stages [Fig.~\ref{fig:forster_gate}]:
first, an effective square $\pi$ pulse ($100\,\mathrm{ns}$) transfers population from $\lvert1\rangle_{\mathrm{Rb}}$ to the Rydberg state $56S_{1/2}$;
second, a shaped optical waveform drives the ytterbium target qubit;
finally, a second phase-coherent $\pi$ pulse de-excites rubidium back to the ground state.
We use rubidium as the control and ytterbium as the target: the longer radiative lifetime of $\mathrm{Rb}(56S)$ ($191.8\,\mu\mathrm{s}$) relative to $\mathrm{Yb}(S)$ ($79.5\,\mu\mathrm{s}$) reduces radiative loss while rubidium idles during the target window.
Accounting for the low-pass control response and ring-down tail (Sec.~\ref{sec:forster_gate_design}), the filtered target window is $160\,\mathrm{ns}$, yielding a total gate duration of $360\,\mathrm{ns}$ ($0.36\,\mu\mathrm{s}$; parameters summarized in Table~\ref{tab:forster_gate}).
Single-qubit dynamical phases accumulated during the sequence are cancelled by virtual local-$Z$ rotations~\cite{McKay2017,Levine2019}.
\end{revnowblock}

The four computational input states evolve as follows:
\begin{itemize}
\item $\lvert 00\rangle$: Both atoms remain decoupled from the optical fields, so the state is unchanged.
\item $\lvert 01\rangle$: Rubidium remains in the ground state, while the shaped Yb pulse drives a single-atom trajectory that returns the population to the computational state with near-unity efficiency and a single-qubit dynamical phase [Fig.~\ref{fig:forster_gate}(c)--(f)].
\item $\lvert 10\rangle$: Ytterbium remains unexcited, while rubidium is promoted to $56S_{1/2}$, idles through the target window, and is coherently returned by the closing $\pi$ pulse, accumulating a single-qubit phase.
\item $\lvert 11\rangle$: Rubidium is promoted to $56S_{1/2}$, so the subsequent Yb drive couples the pair into the strongly interacting F\"orster-dressed manifold.
Dipole--dipole exchange populates both $\lvert SS\rangle$ and $\lvert PP\rangle$, and the shaped waveform returns the population with near-unity efficiency to the control-excited entrance channel. The closing Rb pulse then maps it back to $\lvert 11\rangle$ [Fig.~\ref{fig:forster_gate}(c)--(f)].
\end{itemize}

\begin{figure*}[t]
\centering
\includegraphics[width=\textwidth]{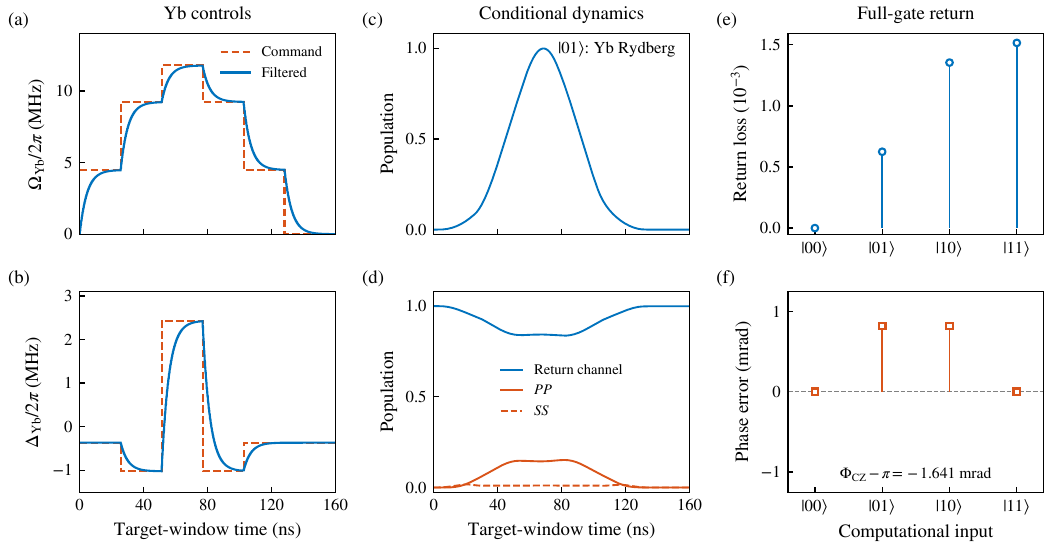}
\caption{\textbf{Pulse control, population return, and the conditional gate phase.}
(a,b)~Yb Rabi frequency and detuning: dashed orange commands and solid blue drives filtered by the assumed first-order response, with \revsym{$t_{\mathrm{r}}=10\,\mathrm{ns}$} (10\%--90\% rise time). Time starts at the target-window entrance and includes the response tail. Two $100\,\mathrm{ns}$ Rb pulses surround this window.
(c,d)~Populations from unnormalized no-jump evolution, initialized separately at the target-window entrance for the unblocked $\lvert01\rangle$ and control-excited $\lvert11\rangle$ branches. The latter visits $SS$ and $PP$ before returning to its entrance channel; the closing Rb pulse maps this channel back to $\lvert11\rangle$.
(e)~Full-sequence return losses $1-|k_{jj}|^2$, including radiative loss and residual excitation.
(f)~Final computational phase errors relative to an ideal $\mathrm{CZ}$ after the nominal local-$Z$ corrections. The conditional phase error is $\approx -1.6\,\mathrm{mrad}$. Both endpoint diagnostics follow from the complex computational return amplitudes $k_{jj}$.
At $R=3.4\,\mu\mathrm{m}$ and $B=3.10\,\mathrm{G}$, the loss-aware average fidelity is $99.91\%$.}
\label{fig:forster_gate}
\end{figure*}

The control--target--control timing follows the sequence of the original Rydberg gate~\cite{Jaksch2000}.
Entanglement is generated by shaping the target waveform so that the exchange-assisted and single-atom trajectories return with different phases, rather than by static blockade.
The relevant gauge invariant is $\Phi_{\mathrm{CZ}}=\arg[k_{00}k_{11}/(k_{01}k_{10})]$, where $k_{ij}$ are the complex computational return amplitudes defined below. Local-$Z$ rotations eliminate single-qubit phases but leave this invariant intact.
The gate requires both population return and conditional phase closure $\Phi_{\mathrm{CZ}}\simeq\pi$.
Homonuclear dipolar schemes use microwave fields to break symmetry and couple opposite-parity levels within identical atoms~\cite{Giudici2025}. In the dual-species $^{87}\mathrm{Rb}$--$^{171}\mathrm{Yb}$ system, species-selective optical transitions directly access the interacting $SS$ state.
The shaped Yb pulse then steers this hybridized manifold through a composite revival without requiring adiabatic following or auxiliary microwave drives.

\subsection{\revnow{Driven Hamiltonian and fidelity metric}}
\label{sec:hamiltonian}
\begin{reveblock}
We model the sequential gate dynamics using conditional no-jump quantum trajectories~\cite{Dalibard1992,Plenio1998}, treating every spontaneous radiative event as absorbing loss outside the computational space.
We use radiative lifetimes and neglect return through decay cascades, blackbody-induced transitions, and motional dynamics.
We propagate state vectors with the time-ordered evolution operator $\mathcal{T}\exp[-2\pi i\int (H(t)/h)\,\mathrm{d}t]$ of the time-dependent Schr\"odinger equation. We express the fields in frequency units, $\bar\Omega(t)=\Omega(t)/(2\pi)$ and $\bar\Delta(t)=\Delta(t)/(2\pi)$.

The opening and closing resonant $\mathrm{Rb}$ $\pi$ pulses are described in the basis $\{\lvert1\rangle_{\mathrm{Rb}},\lvert56S_{1/2}\rangle\}$ by
\begin{equation}
\frac{H_{\mathrm{Rb}}}{h}=
\begin{pmatrix}
0 & \bar\Omega_{\mathrm{Rb}}e^{-i\phi}/2\\
\bar\Omega_{\mathrm{Rb}}e^{i\phi}/2 & -i/(4\pi\tau_{\mathrm{Rb},56S})
\end{pmatrix},
\label{eq:hrbdrive}
\end{equation}
with $\bar\Omega_{\mathrm{Rb}}=5\,\mathrm{MHz}$ and identical optical phase $\phi=0$, ensuring coherent de-excitation of the control atom back to the ground state.

During the target window, the unblocked branch $\lvert01\rangle$ is governed by the two-state Hamiltonian in the ordered basis $\{\lvert01\rangle, \lvert0\rangle_{\mathrm{Rb}}\lvert S\rangle_{\mathrm{Yb}}\}$:
\begin{equation}
\frac{H_{01}(t)}{h}=
\begin{pmatrix}
0 & \bar\Omega(t)/2\\
\bar\Omega(t)/2 & -\bar\Delta(t)-i/(4\pi\tau_{\mathrm{Yb},S})
\end{pmatrix}.
\label{eq:h01}
\end{equation}
For the control-excited branch $\lvert11\rangle$, let \revsym{$\lvert r_c, 1_t\rangle$} $\equiv \lvert56S_{1/2}\rangle_{\mathrm{Rb}}\lvert1\rangle_{\mathrm{Yb}}$ denote the state where rubidium is in $56S$ and ytterbium remains in the computational state.
The optical drive couples \revsym{$\lvert r_c, 1_t\rangle$} to the retained eigenmodes $\lvert\phi_\mu\rangle$ of the full multichannel pair Hamiltonian~\cite{Beterov2015,Beterov2016,Weber2017}, with overlap amplitudes $c_\mu=\langle\phi_\mu\vert SS\rangle$ and relative mode frequencies $\epsilon_\mu=(E_\mu-E_{SS})/h$:
\begin{equation}
\begin{aligned}
\frac{H_{11}(t)}{h}={}&
-\frac{i}{4\pi\tau_{\mathrm{Rb},56S}}\revsym{\lvert r_c, 1_t\rangle\langle r_c, 1_t\rvert}\\
&+\sum_\mu\left[\epsilon_\mu-\bar\Delta(t)-\frac{i\Gamma_\mu}{4\pi}\right]
\lvert\phi_\mu\rangle\langle\phi_\mu\rvert\\
&+\frac{\bar\Omega(t)}{2}\sum_\mu
\left(c_\mu\revsym{\lvert\phi_\mu\rangle\langle r_c, 1_t\rvert}+\mathrm{H.c.}\right),
\end{aligned}
\label{eq:h11}
\end{equation}
where $\mathrm{H.c.}$ denotes the Hermitian conjugate of the optical coupling.
The mode widths $\Gamma_\mu$ are weighted by bare-state composition. This diagonal approximation omits off-diagonal damping elements in the pair-eigenmode basis (Appendix~\ref{app:gate_convergence}); comparison with the full transformed bare-state loss operator gives a branch-loss deviation below $6\times10^{-5}$ throughout the trajectory.

We calibrate the single-qubit phases using virtual local-$Z$ rotations without physical duration or pulse overhead~\cite{McKay2017,Levine2019}:
\begin{equation}
\revsym{V_Z}=\operatorname{diag}(1,e^{i\beta},e^{i\alpha},e^{i(\alpha+\beta)}),
\label{eq:localz}
\end{equation}
where the phases $(\alpha,\beta)$ [Table~\ref{tab:forster_gate}] act on $\mathrm{Rb}$ and $\mathrm{Yb}$, respectively.
We choose them to maximize the nominal loss-aware average fidelity defined below and hold them fixed throughout the position and amplitude tests.
The response-time scan compares fixed and recalibrated phases; the magnetic-field scan tracks the optical carriers and recalibrates the phases [Fig.~\ref{fig:forster_robustness}].
\end{reveblock}

We evaluate gate performance by propagating the complete sequence on the projected multichannel Hamiltonian at the axial, frozen-position baseline ($R_0=3.4\,\mu\mathrm{m}$, $\theta=0$, $B=3.10\,\mathrm{G}$), retaining all electronic $m_J$ sublevels and $^{87}\mathrm{Rb}$ nuclear-spin projections in the axial conserved sector ($M_{\mathrm{tot}}=5/2$, comprising $3684$ states through multipole order $R^{-4}$; Appendix~\ref{app:gate_convergence}).
Here \revsym{$M_{\mathrm{tot}}=m_J^{\mathrm{Rb}}+m_I^{\mathrm{Rb}}+m_F^{\mathrm{Yb}}$} is the total angular-momentum projection along the magnetic axis.
Projection of the no-jump output onto the computational subspace gives the diagonal, trace-decreasing operator $K=\operatorname{diag}(k_{00},k_{01},k_{10},k_{11})$, where $k_{ij}$ is the unnormalized return amplitude for input $\lvert ij\rangle$ and \revsym{$p_{ij}=\lvert k_{ij}\rvert^{2}$} is its computational return probability.
This probability excludes both radiative loss and residual excitation outside the computational subspace.
With $U_{\mathrm{CZ}}=\operatorname{diag}(1,1,1,-1)$, the loss-aware average gate fidelity $F_{\mathrm{avg}}$ is defined as the Haar state average of $\lvert\langle\psi\rvert U_{\mathrm{CZ}}^{\dagger}\revsym{V_Z}K\rvert\psi\rangle\rvert^{2}$ over all two-qubit pure states $\lvert\psi\rangle$, assigning zero target-state overlap to radiative loss and residual excitation.
It is an intrinsic fidelity for this absorbing-loss model, not a trace-preserving treatment of the physical decay channels.
Pedersen, M\o{}ller, and M\o{}lmer~\cite{Pedersen2007} extended the trace-formula framework for average gate fidelity~\cite{Horodecki1999,Nielsen2002} to trace-decreasing quantum operations. Their result gives the uniform Haar average over a $d$-dimensional Hilbert space ($d=4$) as
\begin{equation}
F_{\mathrm{avg}}=\frac{\mathrm{Tr}(K^{\dagger}K)+\lvert\mathrm{Tr}(U_{\mathrm{CZ}}^{\dagger}\revsym{V_Z}K)\rvert^{2}}{d(d+1)},
\label{eq:favg}
\end{equation}
where $d(d+1)=20$.
The first term in the numerator determines the mean computational return probability under no-jump dynamics, $\bar p_{\mathrm{comp}}=\mathrm{Tr}(K^{\dagger}K)/4=\frac{1}{4}\sum_{ij}\lvert k_{ij}\rvert^2$, which sets the upper bound $F_{\mathrm{avg}}\le \bar p_{\mathrm{comp}}$, saturated when \revsym{$V_ZK$} is proportional to $U_{\mathrm{CZ}}$.
The ratio $F_{\mathrm{avg}}/\bar p_{\mathrm{comp}}$ is a return-probability-weighted conditional overlap; it does not measure total leakage or the probability of avoiding a radiative jump.

\begin{table}[tb]
\caption{\textbf{Parameters and performance of the composite F\"orster $\mathrm{CZ}$ gate.}
Control sequence specifications, operating parameters, and loss-aware average gate fidelities $F_{\mathrm{avg}}$ [Eq.~\eqref{eq:favg}] under nominal and bounded error conditions with nominal virtual local-$Z$ phase corrections held fixed.}
\label{tab:forster_gate}
\footnotesize
\begin{ruledtabular}
\begin{tabular}{lc}
Quantity & Value \\
\hline
$R_0$, $\theta$, $B$ & $3.4\,\mu\mathrm{m}$, $0^{\circ}$, $3.10\,\mathrm{G}$ \\
Rb $\pi$ pulse & $5\,\mathrm{MHz}$, $100\,\mathrm{ns}$ \\
Yb command & five segments, $A$--$B$--$C$--$B$--$A$ \\
$A$ $(\Omega,\Delta)/2\pi$ & \reve{$4.48$, $-0.36\,\mathrm{MHz}$} \\
$B$ $(\Omega,\Delta)/2\pi$ & \reve{$9.23$, $-1.02\,\mathrm{MHz}$} \\
$C$ $(\Omega,\Delta)/2\pi$ & \reve{$11.78$, $+2.43\,\mathrm{MHz}$} \\
Segment duration & \reve{$25.6\,\mathrm{ns}$} \\
AOM $10$--$90\%$ response \revsym{($t_{\mathrm{r}}$)} & $10\,\mathrm{ns}$ \\
Filtered target window & \reve{$160\,\mathrm{ns}$} \\
Total gate duration & \reve{$360\,\mathrm{ns}$} \\
Virtual local-$Z$ $(\alpha,\beta)$ & \reve{$(-3.14, 2.94)\,\mathrm{rad}$} \\
\reve{Numerical reference} & \reve{$\pm40\,\mathrm{GHz}$, through $R^{-4}$} \\
\reve{Nominal $F_{\mathrm{avg}}$} & \reve{$99.91\%$} \\
\reve{Mean computational return} & \reve{$99.91\%$} \\
\reve{Conditional subspace overlap} & \reve{$>99.999\%$} \\
\reve{Sampled position-only minimum} & \reve{$99.90\%$} \\
\reve{Sampled joint minimum} & \reve{$99.85\%$} \\
Position bound & $\lvert\delta\mathbf{r}\rvert\le 50\,\mathrm{nm}$ \\
Rabi-scale bounds & Rb/Yb independently $\pm 1\%$ \\
\end{tabular}
\end{ruledtabular}
\end{table}

\subsection{\revnow{Finite-response pulse design and bounded validation}}
\label{sec:forster_gate_design}

\begin{revnowblock}
To represent the finite instrumental bandwidth of the Yb control, the commanded Rabi amplitude and detuning are shaped by an AOM response filter with a $10\text{--}90\%$ rise time \revsym{$t_{\mathrm{r}}=10\,\mathrm{ns}$} and an optical ring-down tail (Appendix~\ref{app:gate_convergence}).
The Yb pulse is parameterized as a five-segment palindromic waveform ($A$--$B$--$C$--$B$--$A$), specified by three independent Rabi frequencies $(\Omega_A,\Omega_B,\Omega_C)$, three detunings $(\Delta_A,\Delta_B,\Delta_C)$, and a uniform segment duration $t_{\mathrm{seg}}=25.6\,\mathrm{ns}$, yielding a filtered target window of $160\,\mathrm{ns}$ (Table~\ref{tab:forster_gate}).
The two Rb pulses remain ideal $100\,\mathrm{ns}$ square pulses in the model.

The filtered waveform brings the computational branches close to a revival [Fig.~\ref{fig:forster_gate}(c)--(f)].
For the unblocked branch $\lvert01\rangle$, the drive excites the Yb Rydberg state and returns the population to the computational state with residual excitation below $4\times10^{-6}$.
For the control-excited branch $\lvert11\rangle$, the drive couples into the hybridized pair manifold, where dipole--dipole exchange populates $\lvert PP\rangle$ (peaking at $15.2\%$) and $\lvert SS\rangle$ (peaking at $1.7\%$), before returning $99.90\%$ of the population to the control-excited entrance channel by the end of the window. The remaining no-jump excitation in the target pair states is $\approx 8.6\times10^{-5}$; radiative loss during this branch contributes $\approx9.1\times10^{-4}$.

Figure~\ref{fig:forster_gate}(e) resolves the full-sequence return loss by computational input; it includes both decay and residual excitation. Figure~\ref{fig:forster_gate}(f) shows the final computational phase errors after the nominal local-$Z$ corrections, obtained from the complex return amplitudes. Their conditional combination gives $\Phi_{\mathrm{CZ}}-\pi\approx -1.6\,\mathrm{mrad}$.
The sensitivity of this same pulse is examined separately in Fig.~\ref{fig:forster_robustness}.

At the reference operating point, the loss-aware average gate fidelity is $F_{\mathrm{avg}}=99.91\%$, matching the mean computational return probability $\bar p_{\mathrm{comp}}=99.91\%$ to within a conditional-overlap error below $2\times10^{-7}$ (conditional subspace overlap $>99.999\%$; Table~\ref{tab:forster_gate}).
The overall infidelity $1-F_{\mathrm{avg}}\approx 8.7\times 10^{-4}$ is overwhelmingly dominated by intrinsic Rydberg radiative decay ($\approx 8.5\times 10^{-4}$).
Coherent control imperfections remain negligible: residual pair excitation at the end of the target window contributes $\approx 2.3\times 10^{-5}$, final spectator leakage stays below $4\times 10^{-7}$ (with a transient maximum of $7.6\times 10^{-4}$), and conditional phase errors ($\Phi_{\mathrm{CZ}}-\pi\approx -1.6\,\mathrm{mrad}$) contribute only $\approx 2\times 10^{-7}$ (detailed budget in Appendix~\ref{app:gate_convergence}, Table~\ref{tab:loss_budget}).

In the axial spectrum, the target bright doublet is well isolated from spectator modes.
The two bright eigenstates are split by $30.7\,\mathrm{MHz}$ (shifted by $-19.0\,\mathrm{MHz}$ and $+11.6\,\mathrm{MHz}$ relative to the bare $\lvert SS\rangle$ asymptote), with a $48.6\,\mathrm{MHz}$ detuning gap to the nearest spectator eigenmode.
Expanding the pair-energy window, including partial $R^{-5}$ multipoles, and refining the numerical integration step shift the nominal fidelity by at most $7.4\times10^{-7}$ (Table~\ref{tab:p04_convergence} and Appendix~\ref{app:gate_convergence}), confirming that numerical truncation errors lie well below the physical loss scales.

\begin{figure}[t]
\centering
\includegraphics[width=\columnwidth]{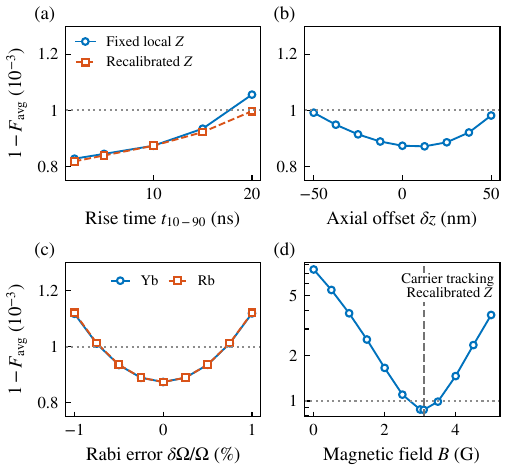}
\caption{\textbf{Sensitivity of the fixed pulse to control and geometry perturbations.}
(a)~Yb response-time scan with local-$Z$ corrections held fixed or recalibrated at each point. The command segments are fixed; the $7\tau$ tail, and hence total duration, varies with response time.
(b,c)~Axial-displacement and individual Rabi-amplitude scans with nominal local-$Z$ corrections held fixed.
All panels express infidelity $1-F_{\mathrm{avg}}$ in units of $10^{-3}$; panels (a)--(c) use linear axes to resolve the narrow range.
(d)~Magnetic-field scan with optical carriers tracking the atomic transitions and local-$Z$ corrections recalibrated at each field, shown on a logarithmic infidelity axis. The vertical line marks $3.10\,\mathrm{G}$.
Dotted horizontal lines mark $1-F_{\mathrm{avg}}=10^{-3}$ throughout. Markers denote calculated points and connecting lines guide the eye.
These one-dimensional scans are distinct from the combined bounded-perturbation validation, whose sampled minimum fidelity is $99.85\%$.}
\label{fig:forster_robustness}
\end{figure}

The final pulse refinement used a regularized minimax objective on nine discrete training configurations: the nominal point and the eight combinations of axial displacement endpoints and independent Rabi-amplitude extremes.
The pulse was then evaluated over a larger set of bounded spatial displacements $\lvert\delta\mathbf{r}\rvert\le 50\,\mathrm{nm}$ and independent $\pm 1\%$ laser amplitude calibration errors [$(s_{\mathrm{Yb}},s_{\mathrm{Rb}})\in\{0.99,1.01\}\times\{0.99,1.01\}$].
One-dimensional parameter sweeps [Fig.~\ref{fig:forster_robustness}(a)--(c)] give the following fidelities near the nominal point:
scanning the AOM rise time from $2\,\mathrm{ns}$ to $20\,\mathrm{ns}$ changes the fidelity from $99.92\%$ to $99.90\%$ (retaining nominal local-$Z$ calibration gives $99.89\%$ at $20\,\mathrm{ns}$),
while isolated scans over axial displacement $\delta z\in[-50,+50]\,\mathrm{nm}$ and $\pm 1\%$ amplitude variations yield sampled minima of $99.90\%$ and $99.89\%$, respectively.

Under simultaneous perturbations across the evaluated uncertainty domain ($\lvert\delta\mathbf{r}\rvert\le 50\,\mathrm{nm}$ and $\pm 1\%$ Rabi scales), the lowest observed fidelity is $99.85\%$, occurring at the axial boundary vertex of $\delta z=-50\,\mathrm{nm}$ ($\theta=0$) where both $\mathrm{Rb}$ and $\mathrm{Yb}$ Rabi frequencies are reduced by $1\%$.
This limiting vertex lies on the symmetry axis where $M_{\mathrm{tot}}=5/2$ is strictly conserved, so it can be evaluated in a closed multichannel hyperfine sector.
For off-axis configurations, dynamics are computed within this retained-symmetry sector as a geometric diagnostic; calculations expanding the pair basis across five total angular-momentum projection sectors ($M_{\mathrm{tot}}=1/2$ to $9/2$, 18,150 states) give cross-sector corrections below $3.5\times 10^{-6}$ over the audited $50\,\mathrm{nm}$ domain (Appendix~\ref{app:gate_convergence}).
Sampling over 37 geometry nodes and dense amplitude grids, followed by surrogate interpolation and adversarial gradient searches, found no lower fidelity than at this axial vertex.
The magnetic-field scan [Fig.~\ref{fig:forster_robustness}(d)] shows an operating range near $B=3.10\,\mathrm{G}$ with infidelities below $10^{-3}$ when optical frequencies track atomic Zeeman shifts and local-$Z$ phases are recalibrated. This scan measures the range available with recalibration, not tolerance to uncompensated magnetic noise.
Systematic uncertainties in atomic level energies require separate calibration; as evaluated in Appendix~\ref{app:spectroscopy_sensitivity}, the fidelity is locally quadratic in the residual pair defect after calibration and remains above $99.84\%$ over $\pm 1\,\mathrm{MHz}$ offsets.
Together, these audits support an exchange-assisted interspecies gate at zero applied electric field, achieving a nominal intrinsic fidelity of $99.91\%$ and a sampled minimum of $99.85\%$ within the modeled error budget.
\end{revnowblock}
\end{revcblock}

\section{\revnow{Discussion and outlook}}
\label{sec:discussion}

\begin{revnowblock}
By identifying a naturally near-resonant zero-electric-field F\"orster doublet ($\Delta_{\mathrm{hfs}}/h \approx -0.69\,\mathrm{MHz}$) with a $30.7\,\mathrm{MHz}$ bright-state splitting at $3.4\,\mu\mathrm{m}$, this protocol circumvents the severe multichannel Stark mixing that afflicts divalent ytterbium under dc bias fields~\cite{Hummel2024,Peper2024}, while species-selective optical addressing bypasses the auxiliary microwave drives needed in homonuclear dipolar gates.
The resulting composite revival $\mathrm{CZ}$ gate operates in $0.36\,\mu\mathrm{s}$ with an intrinsic fidelity of $99.91\%$ ($99.85\%$ sampled minimum under bounded perturbations).
While the auxiliary van der Waals channel ($U/h=57.1\,\mathrm{MHz}$ at $3.3\,\mu\mathrm{m}$, Appendix~\ref{sec:vdw}) offers a complementary static candidate, resonant exchange achieves fast, low-spectator entangling logic at moderate principal quantum numbers.

In the modeled error budget, gate infidelity is governed by Rydberg radiative decay ($\approx 0.085\%$), which exhibits a pronounced species asymmetry: decay of the $^{87}\mathrm{Rb}$ ancilla accounts for $\approx 77\%$ of the total loss ($6.7\times 10^{-4}$), whereas the $^{171}\mathrm{Yb}$ data qubit incurs only $\approx 19\%$ ($1.7\times 10^{-4}$; Appendix~\ref{app:gate_convergence}).
Spontaneous decay from Yb $S$ predominantly leaves the $^{3}P_0$ clock subspace, enabling in-situ optical erasure conversion~\cite{Wu2022,Ma2023,Peper2024}, whereas Rb cascade returns unheralded Pauli errors that the $\mathrm{CZ}$ gate can propagate as correlated phase errors into the data register.
Evaluating the fault-tolerant threshold of this hybrid architecture thus requires embedding these asymmetric channels into full syndrome-extraction circuits~\cite{Zhang2025Dual}.

Translating this benchmark to an experimental realization requires a progressive validation sequence alongside technical error suppression.
Key experimental milestones include:
(i)~single-atom spectroscopy of $^{171}\mathrm{Yb}$ near $302\,\mathrm{nm}$ to pinpoint the Rydberg carrier;
(ii)~interspecies tweezer spectroscopy at $R\approx 3.4\,\mu\mathrm{m}$~\cite{Weber2026} to map the $30.7\,\mathrm{MHz}$ doublet splitting near $B=3.10\,\mathrm{G}$, replacing database inputs with empirical Hamiltonian parameters (Appendix~\ref{app:spectroscopy_sensitivity});
(iii)~time-resolved observation of the predicted $\approx 16\,\mathrm{ns}$ dipole--dipole population swap; and
(iv)~execution of the composite waveform to verify conditional phase closure.
Concurrently, technical intermediate-state scattering during $^{87}\mathrm{Rb}$ excitation~\cite{Saffman2010} is suppressed by large intermediate detunings ($\Delta/2\pi \ge 5\,\mathrm{GHz}$), while trap-induced differential ac Stark shifts can be eliminated by stroboscopic trap switch-off.
\end{revnowblock}

\appendix

\section{Atomic states and pair-interaction model}
\label{app:hyperfine_convergence}

\subsection{Channel-search scope and zero-electric-field selection}
\label{app:channel_search}
The search for an interspecies F\"orster resonance used three stages across candidate $\mathrm{Rb}$--$\mathrm{Yb}$ pair manifolds.
First, a $3\times3$ survey of the $\mathrm{Rb}\,P_{1/2}+\mathrm{Yb}\,P$ manifold around $R=3.5\,\mu\mathrm{m}$ found resonant two-level mixing only for the $56{+}52$ combination; all other combinations exhibited large energy detunings ranging from hundreds of megahertz to gigahertz.
Second, an energy-defect screening over $n\in[45,70]$ for dipole-allowed reaction classes ($P{+}P\to S{+}S$ and $D{+}P\to P{+}D$) found 24 near-degenerate candidate pairs with asymptotic defects $\lvert\delta\rvert/h<30\,\mathrm{MHz}$.
The energy screening depends on the multichannel quantum defect theory (MQDT) parameterization and angular coupling rules. Each near-degenerate candidate must also have appreciable electric dipole matrix elements and limited mixing with background states.
Third, finite-basis diagonalization showed that, in most candidate channels, unwanted state mixing was mediated by strong coupling to isolated perturber states rather than a broad continuum.
Among the examined manifolds, the $\mathrm{Rb}(56S_{1/2})+\mathrm{Yb}(S)\leftrightarrow\mathrm{Rb}(56P_{1/2})+\mathrm{Yb}(P)$ channel was selected for its sub-megahertz asymptotic defect and two-level spectral dominance.
Operating at zero electric field avoids the severe multichannel Stark mixing characteristic of divalent Rydberg systems.
In $^{171}\mathrm{Yb}$, the dense spectrum of opposite-parity Rydberg states causes even modest electric fields to perturb state compositions:
at a field of $0.5\,\mathrm{V/cm}$, the bare target weight of the $\mathrm{Yb}\,P$ partner decreases to $78.6\%$, falling further to $53\%$ at $1.0\,\mathrm{V/cm}$.
Selecting a naturally near-resonant zero-electric-field doublet preserves the target subspace purity without deliberate dc tuning. Gate sensitivity to residual uncompensated stray fields is evaluated in Appendix~\ref{app:bias_fields}.

\subsection{Rubidium hyperfine Hamiltonian and spectroscopic scaling}
\label{app:rb_hyperfine}
In the uncoupled basis \revsym{$\lvert n,L,J,m_J,m_I\rangle$}, the single-atom $^{87}\mathrm{Rb}$ Hamiltonian includes the nuclear spin $I=3/2$:
\begin{equation}
\begin{aligned}
\frac{H_{\mathrm{Rb}}}{h}={}&\revsym{\frac{H_{\mathrm{el}}}{h}}
+\frac{A_{\mathrm{hfs}}}{h}X\\
&+\frac{B_{\mathrm{hfs}}}{h}\mathcal Q_J\\
&+g_I\frac{\mu_B}{h}B_zI_z,
\qquad X=\mathbf I\!\cdot\!\mathbf J.
\end{aligned}
\label{eq:rb_hfs}
\end{equation}
The dimensionless quadrupole operator vanishes for $J=1/2$. For $J\ge1$,
\begin{equation}
\mathcal Q_J=\frac{3X^2+\tfrac32X-I(I+1)J(J+1)}{2I(2I-1)J(2J-1)}.
\end{equation}
\revsym{$H_{\mathrm{el}}$} is the uncoupled electronic Hamiltonian evaluated by PairInteraction and includes fine structure and the electronic Zeeman interaction. The nuclear $g$-factor is $g_I=-9.951414\times10^{-4}$.

For the Rb hyperfine scaling, $n^*=n-\delta_{\ell j}(n)$ is the effective principal quantum number, with the modified Rydberg--Ritz convention $\delta_{\ell j}(n)=\delta_0+\delta_2/(n-\delta_0)^2$ and series-dependent coefficients.
The magnetic-dipole constants $A_{\mathrm{hfs}}$ and electric-quadrupole constants $B_{\mathrm{hfs}}$ for high Rydberg levels are determined through $(n^*)^{-3}$ scaling of experimentally established low-lying states:
(i)~For $56S_{1/2}$, we adopt the measured scaling $A/h=18.55\,\mathrm{GHz}/(n^*)^3$~\cite{Tauschinsky2013}, yielding $A_{56S}/h=125.5\,\mathrm{kHz}$ and a zero-field $F=2$--$F=1$ hyperfine splitting of $251.1\,\mathrm{kHz}$.
(ii)~For $56P_{1/2}$, isotope-scaling the measured $^{85}\mathrm{Rb}$ normalization by the nuclear $g$-factor ratio yields $A_{56P_{1/2}}/h=32.2\,\mathrm{kHz}$ and an unperturbed splitting of $64.4\,\mathrm{kHz}$~\cite{Cardman2022}.
(iii)~Couplings to low-lying $P_{3/2}$ and $D_J$ manifolds are incorporated via $(n^*)^{-3}$ scaling fits to compiled spectroscopic data~\cite{Arimondo1977}.
Because experimental hyperfine normalizations are unavailable for weakly coupled $F_J$ spectator states, their central hyperfine constants are set to zero in the reference basis.
Numerically assigning these spectator states the largest fitted low-$\ell$ normalizations alters the phase-recalibrated gate fidelity by less than $10^{-8}$, showing negligible sensitivity to these spectator splittings in the tested model.

\subsection{State specification, MQDT conventions, and working-state dictionary}
\label{app:state_dictionary}
\begin{table*}[t]
\caption{\textbf{Working-state dictionary and Rydberg level parameters.} The Rb nuclear projection is retained explicitly in the F\"orster gate.}
\label{tab:state_dictionary}
\scriptsize
\newcommand{\wcol}[2]{\parbox[t]{#1\textwidth}{\raggedright\hspace{0pt}#2}}
\begin{ruledtabular}
\begin{tabular}{llll}
\wcol{0.12}{Role} & \wcol{0.27}{Physical state or transition} & \wcol{0.22}{Selected magnetic component} & \wcol{0.30}{Database selector and model input} \\
\hline
\wcol{0.12}{Rb F\"orster drive}
& \wcol{0.27}{$5S_{1/2},F{=}2,m_F{=}+2\rightarrow56S_{1/2}$}
& \wcol{0.22}{$m_J{=}+1/2$, $m_I{=}+3/2$}
& \wcol{0.30}{$\mathrm{Rb}(56,0,1/2,+1/2)$; $g_J=2.0023$; $\tau_0=191.8\,\mu\mathrm{s}$; $5\,\mathrm{MHz}$ effective drive} \\
\wcol{0.12}{Yb F\"orster drive}
& \wcol{0.27}{$^{3}P_0,F{=}1/2,m_F{=}-1/2\rightarrow S(\nu{=}48.369927)$}
& \wcol{0.22}{$F{=}1/2$, $m_F{=}+1/2$}
& \wcol{0.30}{$(53,0,1,1/2,+1/2)$; $g_F=2.4414$; $\tau_0=79.5\,\mu\mathrm{s}$; $\lvert d_{+1}\rvert=2.39\!\times\!10^{-3}ea_0$} \\
\wcol{0.12}{F\"orster $PP$ partner}
& \wcol{0.27}{$\mathrm{Rb}\,56P_{1/2}+\mathrm{Yb}\,P(\nu{=}48.014048)$}
& \wcol{0.22}{$m_J{=}+1/2,m_I{=}+3/2$; $F{=}1/2,m_F{=}+1/2$}
& \wcol{0.30}{$\mathrm{Rb}(56,1,1/2,+1/2)$, $g_J=0.6659$; Yb $(52,1,0,1/2,+1/2)$, $g_F=1.3112$; $\tau_0=(413.8,340.7)\,\mu\mathrm{s}$} \\
\wcol{0.12}{Rb vdW candidate}
& \wcol{0.27}{$\mathrm{Rb}\,66S_{1/2}$}
& \wcol{0.22}{$m_J{=}+1/2$ (stretched sector)}
& \wcol{0.30}{$\mathrm{Rb}(66,0,1/2,+1/2)$; $g_J=2.0023$; $\tau_0=323.6\,\mu\mathrm{s}$} \\
\wcol{0.12}{Yb vdW candidate}
& \wcol{0.27}{$S(\nu{=}62.682293)$, $E=50415.288\,\mathrm{cm}^{-1}$}
& \wcol{0.22}{$F{=}1/2$, $m_F{=}+1/2$}
& \wcol{0.30}{$(67,0,0,1/2,+1/2)$; $g_F=0.4007$; $\tau_0=87.6\,\mu\mathrm{s}$; $\lvert d_0\rvert=6.27\!\times\!10^{-4}ea_0$} \\
\end{tabular}
\end{ruledtabular}
\end{table*}
Because integer principal quantum numbers and spin labels $(n, s)$ can shift across different multichannel quantum defect parameterizations and database releases, we identify each ytterbium state by its excitation energy, effective quantum number $\nu$, and spectroscopic term symbol.
In the $\mathrm{Yb171\_mqdt}$ database v1.4~\cite{Peper2024,Kuroda2025}, the target ytterbium levels correspond to selectors $(n{=}53, \ell{=}0, s{=}1, f{=}1/2, m_F{=}+1/2)$ for $\mathrm{Yb}\,S(\nu{=}48.369927)$ and $(n{=}52, \ell{=}1, s{=}0, f{=}1/2, m_F{=}+1/2)$ for $\mathrm{Yb}\,P(\nu{=}48.014048)$.

Table~\ref{tab:state_dictionary} summarizes the state definitions, database selectors, effective linear Zeeman factors $g_J$ or $g_F$, radiative lifetimes $\tau_0$, and optical coupling parameters used across the F\"orster gate and auxiliary van der Waals calculations.
All atomic energies, matrix elements, and pair interactions are evaluated using PairInteraction 2.5.0~\cite{Mogerle2026,Weber2017} with databases $\mathrm{Rb}$ v1.2~\cite{Li2003,Mack2011}, $\mathrm{Yb171\_mqdt}$ v1.4~\cite{Peper2024,Kuroda2025}, and $\mathrm{misc}$ v1.4; database asset hashes in the simulation manifests identify the numerical inputs used in these calculations.
The many digits in the state selectors identify database roots and should not be interpreted as experimental energy uncertainties.
The electronic centroid defect $\Delta_{\mathrm{el}}/h = -0.76\,\mathrm{MHz}$ ($-0.764\,\mathrm{MHz}$ before rounding) evaluated by PairInteraction omits the $^{87}\mathrm{Rb}$ nuclear spin; incorporating the hyperfine Hamiltonian of Eq.~\eqref{eq:rb_hfs} for the stretched states yields the net defect $\Delta_{\mathrm{hfs}}/h = -0.69\,\mathrm{MHz}$ ($-0.694\,\mathrm{MHz}$) given by Eq.~\eqref{eq:defect}.

\subsection{Analytical two-state exchange model and multichannel unitary bound}
\label{app:two_state_bound}
For an analytical reference, we consider the projected two-state subspace $\{\lvert PP\rangle,\lvert SS\rangle\}$ with asymptotic detuning $\delta=\Delta_{\mathrm{el}}$.
Under the dipole--dipole coupling $V=\langle PP\rvert V_{\mathrm{dd}}\lvert SS\rangle$, the effective two-state Hamiltonian and generalized oscillation frequency are
\begin{equation}
\frac{H_2}{h}=
\begin{pmatrix}
0 & V/h\\
V^*/h & \delta/h
\end{pmatrix},\qquad
\Delta\nu_2=\sqrt{(\delta/h)^2+4\lvert V/h\rvert^2}.
\label{eq:two_state_forster}
\end{equation}
Starting from $\lvert SS\rangle$, the dynamic population transfer into $\lvert PP\rangle$ is $P(t) = P_{\max}^{(2)}\sin^2(\pi\Delta\nu_2 t)$, with the peak efficiency and first half-period given by
\begin{equation}
P_{\max}^{(2)}=\frac{4\lvert V\rvert^2}{\delta^2+4\lvert V\rvert^2},
\qquad t_{\max}^{(2)}=\frac{1}{2\Delta\nu_2}.
\label{eq:detuned_exchange}
\end{equation}
At zero detuning ($\delta=0$), these reduce to the resonant energy splitting $\Delta E=2\lvert V\rvert$ and transfer time $t_{\max}=h/(4\lvert V\rvert)$.
At $R=3.4\,\mu\mathrm{m}$, the calculated matrix element $\lvert V\rvert/h=15.4\,\mathrm{MHz}$ yields $\Delta\nu_2=30.8\,\mathrm{MHz}$, $P_{\max}^{(2)}=99.94\%$, and $t_{\max}^{(2)}\approx 16.2\,\mathrm{ns}$.

In the full multichannel Hilbert space, unperturbed unitary evolution from $\lvert SS\rangle$ expands as $\lvert\psi(t)\rangle = \sum_k c_k e^{-i E_k t/\hbar} \lvert\phi_k\rangle$ with $c_k = \langle\phi_k\rvert SS\rangle$.
The time-dependent transfer amplitude into $\lvert PP\rangle$ satisfies the triangle inequality:
\begin{equation}
\lvert\langle PP\rvert\psi(t)\rangle\rvert \le \sum_k \lvert\langle PP\rvert\phi_k\rangle\langle\phi_k\rvert SS\rangle\rvert.
\label{eq:unitary_bound}
\end{equation}
Partitioning the eigenbasis into the bright-state doublet ($k \in \{\pm\}$) and the orthogonal subspace of off-resonant spectator states ($k \in \mathrm{spec}$), we apply the Cauchy--Schwarz inequality to the spectator sum:
\begin{equation}
\begin{aligned}
\sum_{k \in \mathrm{spec}} &\lvert\langle PP\rvert\phi_k\rangle\langle\phi_k\rvert SS\rangle\rvert \\
&\le \sqrt{\sum_{k \in \mathrm{spec}} \lvert\langle PP\rvert\phi_k\rangle\rvert^2} \, \sqrt{\sum_{k \in \mathrm{spec}} \lvert\langle SS\rvert\phi_k\rangle\rvert^2}.
\end{aligned}
\label{eq:cauchy_schwarz}
\end{equation}
For the two bright eigenstates at $R=3.4\,\mu\mathrm{m}$ [Fig.~\ref{fig:forster_channel}(b)], the direct projection contribution from the rounded weights is $\sqrt{0.57755 \times 0.42099} + \sqrt{0.42168 \times 0.57674} \simeq 0.98625$.
The spectator weights are \revsym{$w_{\mathrm{sp}}^{PP} = 0.077\%$} and \revsym{$w_{\mathrm{sp}}^{SS} = 0.227\%$}, yielding a Cauchy--Schwarz bound \revsym{$\sqrt{w_{\mathrm{sp}}^{PP} w_{\mathrm{sp}}^{SS}} \simeq 0.00132$}.
Combining Eqs.~\eqref{eq:unitary_bound} and \eqref{eq:cauchy_schwarz} establishes a phase-independent unitary upper bound on the transfer amplitude: $\lvert\langle PP\rvert\psi(t)\rangle\rvert \le 0.98625 + 0.00132 \simeq 0.98757$, corresponding to a loose population ceiling of $(0.98757)^2 \approx 97.53\%$.
Directly summing the mode-by-mode overlap moduli across the discrete multichannel spectrum tightens this static ceiling to $97.31\%$.
The dynamically propagated peak transfer of $97.29\%$ closely approaches this theoretical bound, consistent with exchange dominated by the bright-state doublet and sub-percent spectator participation.

\subsection{Spectroscopic basis convergence and angular sensitivity}
\label{app:spec_convergence}
We tested basis convergence one parameter at a time for the zero-field fixed-$m$ spectroscopic model [Figs.~\ref{fig:forster_channel}(a)--\ref{fig:forster_channel}(d)] at $R=3.4\,\mu\mathrm{m}$.
The reference spectroscopic basis retains $\Delta n=3$, $\ell\le 3$, and a $\pm80\,\mathrm{GHz}$ pair-energy window under dipole--dipole coupling.
Expanding the pair-energy window to $\pm100\,\mathrm{GHz}$, increasing the principal quantum number cutoff to $\Delta n=4$, or extending the orbital cutoff to $\ell_{\max}=4$ shifts the bright-state splitting by at most $0.00074\,\mathrm{MHz}$, the dynamic transfer efficiency by $0.0045\%$, and spectator population by $0.0023\%$.
The splitting changes by less than $10^{-3}\,\mathrm{MHz}$ in these one-parameter basis expansions; the population changes quoted above quantify convergence within the reference atomic model.

Figure~\ref{fig:forster_channel}(f) examines the angular sensitivity of the exchange interaction in this fixed-$m$ framework.
The primary $q=0$ dipole--dipole tensor component scales as $(1-3\cos^2\theta)$, which vanishes at the magic angle $\theta=\arccos(1/\sqrt3)\simeq54.74^\circ$, while $q=\pm1,\pm2$ components couple distinct magnetic sectors.
In the projected symmetry sector, the scan shows how geometry changes the $q=0$ coupling relative to the fixed defect $\lvert\delta\rvert/h=0.76\,\mathrm{MHz}$.
The sampled dip at $\theta=60^\circ$ and recovery at $75^\circ$ follow this angular dependence: at these orientations, the effective coupling $\lvert V\rvert/h$ evaluates to $1.93\,\mathrm{MHz}$ and $6.15\,\mathrm{MHz}$, yielding two-state peak transfers of $96.2\%$ and $99.6\%$, respectively, compared with full multichannel values of $95.6\%$ and $98.6\%$.
Aligning the interatomic vector along the quantization axis ($\theta=0$) maximizes the dominant $q=0$ exchange strength. Appendix~\ref{app:gate_convergence} evaluates the driven gate under transverse displacements, using both the retained-symmetry model and multi-sector calculations that include $q=\pm1,\pm2$ couplings.

\section{Pulse construction and numerical validation}
\label{app:gate_convergence}

\subsection{Multichannel basis convergence, spectator dynamics, and numerical stability}
\begin{table*}[t]
\caption{\textbf{Numerical convergence of the selected fixed F\"orster-gate pulse.}
The numerical reference (Ref.) uses $\Delta n=3$, $\ell_{\max}=3$, a $\pm80\,\mathrm{GHz}$ atomic window, a $\pm40\,\mathrm{GHz}$ pair window, and interactions complete through $R^{-4}$.
Each other row varies one setting.
$N$ is the connected axial hyperfine block, $w_{\min}$ is the smaller $SS{+}PP$ weight of the two target modes, \revsym{$P_{\mathrm{swap}}^{\max}$} is the maximum static $PP\to SS$ transfer, and \revsym{$P_{\mathrm{spec}}^{\max}$} is the largest spectator population during the fixed driven pulse.
$F_{\mathrm{fix}}$ uses the Ref. local-$Z$ phases; $F_{Z}$ recalibrates only those phases for the row.
Full precision and the cutoff/time-step checks are in the machine-readable record.}
\label{tab:p04_convergence}
\scriptsize
\begin{ruledtabular}
\begin{tabular}{lrrrrrrr}
One-at-a-time setting & $N$ & $\Delta\nu$ (MHz) & $w_{\min}$ (\%) & \revsym{$P_{\mathrm{swap}}^{\max}$} (\%) & \revsym{$P_{\mathrm{spec}}^{\max}$} (\%) & $F_{\mathrm{fix}}$ (\%) & $F_Z$ (\%) \\
\hline
$\pm20\,\mathrm{GHz}$ pair window & 2075 & 30.72486 & 99.54846 & 98.32421 & \revnow{0.07387} & \revnow{99.89115} & \revnow{99.90540} \\
Ref.: $\pm40\,\mathrm{GHz}$, through $R^{-4}$ & 3684 & 30.64741 & 99.54777 & 98.81177 & \revnow{0.07619} & \revnow{99.91258} & \revnow{99.91258} \\
$\pm60\,\mathrm{GHz}$ pair window & 5212 & 30.65471 & 99.54692 & 98.79627 & \revnow{0.07651} & \revnow{99.91250} & \revnow{99.91251} \\
$\Delta n=4$ & 4658 & 30.64262 & 99.54653 & 98.79206 & \revnow{0.07648} & \revnow{99.91255} & \revnow{99.91261} \\
$\pm160\,\mathrm{GHz}$ atomic window & 6359 & 30.64744 & 99.54776 & 98.81151 & \revnow{0.07620} & \revnow{99.91258} & \revnow{99.91258} \\
$\ell_{\max}=4$ & 4398 & 30.64741 & 99.54777 & 98.81178 & \revnow{0.07619} & \revnow{99.91258} & \revnow{99.91258} \\
dipole--dipole only & 1872 & 30.78321 & 99.74761 & 98.84960 & \revnow{0.01949} & \revnow{99.87899} & \revnow{99.90193} \\
partial order $R^{-5}$ & 3684 & 30.66929 & 99.54822 & 98.81088 & \revnow{0.07591} & \revnow{99.91252} & \revnow{99.91252} \\
\end{tabular}
\end{ruledtabular}
\end{table*}
To test numerical convergence of the composite $\mathrm{CZ}$ gate, we expanded the multichannel basis beyond the reference model.
The reference model includes all fine-structure and $^{87}\mathrm{Rb}$ hyperfine states within a single-atom energy window of $\pm80\,\mathrm{GHz}$, a pair-energy window of $\pm40\,\mathrm{GHz}$, radial cutoffs $\Delta n=3$, orbital angular momentum cutoff $\ell_{\max}=3$, and electrostatic multipole interactions complete through $R^{-4}$ (incorporating dipole--dipole, dipole--quadrupole, and quadrupole--dipole couplings).
In the axial conserved symmetry sector ($M_{\mathrm{tot}}=5/2$), the resulting connected Hamiltonian block comprises $N=3684$ states.
At the operating magnetic field $B=3.10\,\mathrm{G}$, differential Zeeman tuning shifts the field-dressed asymptotic Förster defect between stretched states to $+4.66\,\mathrm{MHz}$ ($4.6639\,\mathrm{MHz}$ in the raw model).
The two target bright eigenstates lie at $-19.0\,\mathrm{MHz}$ and $+11.6\,\mathrm{MHz}$ relative to the bare $\lvert SS\rangle$ asymptote (splitting $30.65\,\mathrm{MHz}$).

We include spontaneous radiative decay through non-Hermitian no-jump evolution, with no return of lost population to the computational subspace.
For each coupled eigenmode $\lvert\phi_\mu\rangle$, the mode-specific decay width is constructed as
\begin{equation}
\begin{aligned}
\Gamma_\mu={}&w_{\mu,PP}\Gamma_{PP}+w_{\mu,SS}\Gamma_{SS}\\
&+\revsym{(1-w_{\mu,PP}-w_{\mu,SS})}\max(\Gamma_{PP},\Gamma_{SS}),
\end{aligned}
\label{eq:gamma_mu}
\end{equation}
where $w_{\mu,j}=\lvert\langle j\vert\phi_\mu\rangle\rvert^2$ is the bare-state probability weight of mode $\lvert\phi_\mu\rangle$, \revsym{$1-w_{\mu,PP}-w_{\mu,SS}$} accounts for residual spectator admixture, and composite bare decay widths are $\Gamma_{PP}=\tau_{\mathrm{Rb},56P}^{-1}+\tau_{\mathrm{Yb},P}^{-1}$ and $\Gamma_{SS}=\tau_{\mathrm{Rb},56S}^{-1}+\tau_{\mathrm{Yb},S}^{-1}$.
Equation~\eqref{eq:gamma_mu} retains mode-specific diagonal decay widths in the pair-eigenmode basis.
Transforming bare-state loss operators into this hybridized basis produces off-diagonal damping elements that are omitted here; as noted in Sec.~\ref{sec:hamiltonian}, direct integration gives branch-loss deviations below $6\times10^{-5}$ across the trajectory.
The prescription $\max(\Gamma_{PP},\Gamma_{SS})$ assigns the bare decay rate of the most rapidly decaying target partner ($\Gamma_{SS}$, dominated by $\mathrm{Yb}(S)$ with $\tau=79.5\,\mu\mathrm{s}$) to weakly coupled spectator admixtures. Because all relevant background Rydberg levels within the $\pm40\,\mathrm{GHz}$ pair window have effective principal quantum numbers $n^*\gtrsim 48$ with radiative lifetimes comparable to or longer than $\mathrm{Yb}(S)$ (typically hundreds of microseconds for high-$\ell$ states), this prescription provides a conservative upper damping scale for background radiative loss.
Calculated radiative lifetimes from PairInteraction are $\tau_{\mathrm{Rb},56S}=191.8\,\mu\mathrm{s}$, \revsym{$\tau_{\mathrm{Rb},56P}=413.8\,\mu\mathrm{s}$}, $\tau_{\mathrm{Yb},S}=79.5\,\mu\mathrm{s}$, and $\tau_{\mathrm{Yb},P}=340.7\,\mu\mathrm{s}$, as listed in the working-state dictionary (Table~\ref{tab:state_dictionary}).

Table~\ref{tab:p04_convergence} reports the changes under multichannel basis expansion.
Extending the pair-energy window to $\pm60\,\mathrm{GHz}$, enlarging the single-atom window to $\pm160\,\mathrm{GHz}$, or extending cutoffs to $\Delta n=4$ and $\ell_{\max}=4$ shifts the fixed-phase gate fidelity by at most $7.4\times10^{-7}$.
Incorporating leading quadrupole--quadrupole interactions (partial order $R^{-5}$) changes the fidelity by $5.5\times10^{-7}$, relative to the complete $R^{-4}$ baseline.
We also tested the pair-eigenmode projection cutoff and time step: lowering the mode-retention threshold from $10^{-6}$ to $10^{-8}$ shifts the nominal fidelity by $4.4\times10^{-11}$, while refining the integration time step from $0.125\,\mathrm{ns}$ to $0.03125\,\mathrm{ns}$ induces a shift of $9.55\times10^{-10}$.
Transient non-adiabatic spectator dynamics are strongly suppressed by the $48.6\,\mathrm{MHz}$ spectral detuning gap to the nearest spectator mode:
total population in all spectator states peaks at $7.62\times10^{-4}$ during the drive and returns to $3.86\times10^{-7}$ at the target-window endpoint, so the final spectator population is small compared with the computational return loss.

\begin{table}[tb]
\caption{\textbf{Computational return loss and infidelity budget at the reference operating point.}
Full-sequence computational return losses $1-\lvert k_{ij}\rvert^2$ and their contributions to the mean return loss $\frac{1}{4}(1-\lvert k_{ij}\rvert^2)$ across input states $\lvert ij\rangle$, alongside the conditional-overlap defect and nominal gate infidelity.}
\label{tab:loss_budget}
\scriptsize
\begin{ruledtabular}
\begin{tabular}{lcc}
Branch $\lvert ij\rangle$ (trajectory) & $1-\lvert k_{ij}\rvert^2$ & Contribution \\
\hline
$\lvert00\rangle$ (idle ground) & $0.0000$ & $0.0000$ \\
$\lvert01\rangle$ (Yb $2\pi$, $49.4\,\mathrm{ns}$ Rydberg) & $6.2489\times 10^{-4}$ & $1.5622\times 10^{-4}$ \\
$\lvert10\rangle$ (Rb $\pi$--idle--$\pi$) & $1.3550\times 10^{-3}$ & $3.3875\times 10^{-4}$ \\
$\lvert11\rangle$ (Rb $\pi$--pair--$\pi$) & $1.5161\times 10^{-3}$ & $3.7904\times 10^{-4}$ \\
\hline
Mean return loss $1-\bar p_{\mathrm{comp}}$ & \multicolumn{2}{c}{$8.7401\times 10^{-4}$} \\
Conditional defect $\bar p_{\mathrm{comp}}-F_{\mathrm{avg}}$ & \multicolumn{2}{c}{$2.08\times 10^{-7}$} \\
Gate infidelity $1-F_{\mathrm{avg}}$ ($99.91\%$) & \multicolumn{2}{c}{$8.7422\times 10^{-4}$} \\
\end{tabular}
\end{ruledtabular}
\end{table}

Table~\ref{tab:loss_budget} lists the branch-resolved computational return losses $1-\lvert k_{ij}\rvert^2$ obtained from the full-sequence no-jump propagator $K$.
The contributions to the gate infidelity are:
(i) Rydberg radiative decay accounts for $\approx 8.52\times 10^{-4}$ of the infidelity, mainly from the $^{87}\mathrm{Rb}(56S)$ ancilla ($\approx 6.7\times 10^{-4}$, reflecting $260\,\mathrm{ns}$ integrated residence time in $\lvert10\rangle$ alongside control pulses in $\lvert11\rangle$), while $^{171}\mathrm{Yb}$ decay contributes $\approx 1.7\times 10^{-4}$ (governed by $49.4\,\mathrm{ns}$ integrated Rydberg population during the unblocked drive);
(ii) Incomplete population return (residual non-radiative pair excitation) at the end of the driven target window contributes $\approx 2.25\times 10^{-5}$ (dominated by the $8.63\times 10^{-5}$ target-window pair residual in $\lvert11\rangle$); and
(iii) Coherent phase and conditional-overlap defects contribute $\bar p_{\mathrm{comp}}-F_{\mathrm{avg}} = 2.08\times 10^{-7}$.

\subsection{Pulse parameterization and sampled robustness}
The five-segment palindromic target waveform ($A$--$B$--$C$--$B$--$A$) introduced in Sec.~\ref{sec:forster_gate_design} is defined by seven independent control parameters: three Rabi frequencies $(\Omega_A, \Omega_B, \Omega_C)$, three detunings $(\Delta_A, \Delta_B, \Delta_C)$, and a uniform segment duration $t_{\mathrm{seg}}$.
To model the finite response of the AOM and RF drive electronics, commanded waveforms pass through first-order low-pass filters with a $10\text{--}90\%$ rise time \revsym{$t_{\mathrm{r}}=10\,\mathrm{ns}$}:
\begin{equation}
\tau=\frac{\revsym{t_{\mathrm{r}}}}{\ln 9}\approx 4.55\,\mathrm{ns},
\label{eq:aom_tau}
\end{equation}
governing the instantaneous parameters via
\begin{equation}
\dot\Omega=\frac{\Omega_{\mathrm{cmd}}-\Omega}{\tau},\qquad
\dot\Delta=\frac{\Delta_{\mathrm{cmd}}-\Delta}{\tau}.
\label{eq:aom}
\end{equation}
Following the commanded segments, an optical ring-down tail of duration $7\tau\approx 31.9\,\mathrm{ns}$ ($31.86\,\mathrm{ns}$) is appended, during which the amplitude command is zero and the detuning command is held at its final value.
The filtered detuning continues to relax toward that value, and the amplitude is truncated after falling by a factor $e^{-7}$.
The filtered amplitude starts at zero and the detuning starts at its first command value.
The command sequence is palindromic, but its causal filtered response need not be time symmetric.

To test sensitivity to position and Rabi-amplitude errors, we consider bounded spatial displacements and laser-scaling errors:
\begin{equation}
\mathbf{R}=R_0\hat{\mathbf{z}}+\delta\mathbf{r},\qquad \lvert\delta\mathbf{r}\rvert\le 50\,\mathrm{nm},
\label{eq:pos_ball}
\end{equation}
with independent $\pm 1\%$ laser amplitude calibration errors spanning the amplitude vertices $(s_{\mathrm{Yb}}, s_{\mathrm{Rb}})\in\{0.99, 1.01\}\times\{0.99, 1.01\}$.
For the final reference-basis refinement, the discrete training set $\mathcal T$ contains nine configurations: the nominal operating point and the eight combinations of axial displacement endpoints $\delta z=\pm50\,\mathrm{nm}$ and independent laser amplitude extremes [$(s_{\mathrm{Yb}}, s_{\mathrm{Rb}})\in\{0.99, 1.01\}\times\{0.99, 1.01\}$].
The refinement minimizes the regularized minimax objective:
\begin{equation}
J=\max_{x\in\mathcal T}[1-F_{\mathrm{avg}}(x)]
+\frac{0.02}{\lvert\mathcal T\rvert}\sum_{x\in\mathcal T}[1-F_{\mathrm{avg}}(x)],
\label{eq:objective}
\end{equation}
where the first term is the worst-case training-set infidelity and the second term (weight $0.02$) penalizes the mean infidelity.
Following initial seven-parameter exploration, deterministic Nelder--Mead refinement varied the six amplitude and detuning parameters within box constraints ($0\le\Omega_i/2\pi\le15\,\mathrm{MHz}$, $-5\le\Delta_i/2\pi\le5\,\mathrm{MHz}$) at fixed segment duration $t_{\mathrm{seg}}\approx 25.6\,\mathrm{ns}$ ($25.634\,\mathrm{ns}$), terminating at its 400-iteration ceiling (625 objective evaluations).
The resulting pulse was tested under the perturbations described below; its parameters are listed in Table~\ref{tab:forster_gate}.
All subsequent evaluations include $^{87}\mathrm{Rb}(56S)$ radiative decay across the control-excited target window.

We tested the fixed pulse in four ways:
\begin{enumerate}
\item \textit{Boundary and spherical shell mapping}: Performance was evaluated across 19 spatial configurations (origin plus two spherical shells of radii $25\,\mathrm{nm}$ and $50\,\mathrm{nm}$ sampled at nine direction cosines) combined with the four amplitude extremes. The spatial-only minimum fidelity is $99.90\%$ ($99.9008\%$ in the numerical record), and the sampled joint minimum is $99.85\%$ ($99.8499\%$), occurring at the negative axial boundary $\delta z=-50\,\mathrm{nm}$ ($\theta=0$) with both Rabi frequencies scaled by $-1\%$. Because this vertex lies on the symmetry axis, $M_{\mathrm{tot}}=5/2$ is conserved and there is no angular projection error. To test off-axis sensitivity to broken cylindrical symmetry, we expanded the frozen-position basis to five projection sectors ($M_{\mathrm{tot}}=1/2,\dots,9/2$) at representative off-axis nodes. The most negative fidelity shift was $-3.5\times 10^{-6}$ relative to the retained-symmetry projection (with transient cross-sector population peaking below $6.3\times 10^{-5}$), consistent with weak contributions from $q=\pm1,\pm2$ dipolar couplings at these detunings.
\item \textit{Dense multi-parameter grid}: Direct reference-model evaluation over 37 geometry nodes (four radial shells up to $50\,\mathrm{nm}$ at nine direction cosines) combined with a 25-point amplitude grid spanning $\pm1\%$ gave the same sampled minimum of $99.85\%$ ($99.8499\%$).
\item \textit{Surrogate response surface}: A cubic tensor-product interpolant constructed over the evaluation grid was validated against six independent Hamiltonian-evaluated points, yielding maximum and rms holdout discrepancies of $1.57\times10^{-5}$ and $9.75\times10^{-6}$. Probing the interpolant with nested Sobol sequences up to 1024 points found no lower fidelity in the interior of the sampled domain.
\item \textit{Adversarial gradient search}: Local SLSQP searches on the surrogate surface initialized from the 12 lowest Sobol nodes yielded candidate local minima. Direct Hamiltonian re-evaluation of distinct candidate configurations found no fidelity below that at the negative axial grid endpoint.
\end{enumerate}
These checks give $99.85\%$ ($99.8499\%$ in the numerical record) as the sampled minimum gate fidelity observed across the evaluated perturbation grids, Sobol designs, and local adversarial searches.

\subsection{Thermal-motion scales and the tested geometry domain}
\label{app:thermal_motion}
The nominal benchmark fidelity of $99.91\%$ and its sampled minimum of $99.85\%$ refer to the axial, frozen-position reference model ($R_0=3.4\,\mu\mathrm{m}$, $\theta=0$), where $M_{\mathrm{tot}}=5/2$ is conserved.
To test sensitivity to trap misalignment and drift, we evaluated performance over the deterministic bound $\lvert\delta\mathbf{r}\rvert\le 50\,\mathrm{nm}$ within the retained-symmetry sector.
To relate this bound to physical thermal scales, we consider representative tweezer parameters: $T=2.9\,\mu\mathrm{K}$ with trap frequencies $(60,60,10)\,\mathrm{kHz}$ for $^{171}\mathrm{Yb}$~\cite{Peper2024}, and $T=13\,\mu\mathrm{K}$ with $(154,150,30)\,\mathrm{kHz}$ for $^{87}\mathrm{Rb}$~\cite{Kaufman2012}.
Within the tightly confined transverse focal plane of typical two-dimensional tweezer arrays, these parameters yield relative root-mean-square thermal position widths of $(\sigma_x,\sigma_y) \approx (49.8,50.5)\,\mathrm{nm}$, comparable to the displacement scale tested here.
Motion along the weakly confined tweezer propagation axis ($\sigma_\parallel \approx 266\,\mathrm{nm}$) is perpendicular to the interatomic quantization axis $\hat{\mathbf{z}}$ in planar array architectures, shifting the interatomic separation only to second order:
\begin{equation}
\Delta R = \sqrt{R_0^2 + \sigma_\parallel^2} - R_0 \simeq \frac{\sigma_\parallel^2}{2R_0} \approx 10.4\,\mathrm{nm} \ll 50\,\mathrm{nm},
\end{equation}
accompanied by a geometric out-of-plane tilt of $\theta \simeq \sigma_\parallel / R_0 \approx 4.5^\circ$.
The corresponding single-photon ($302\,\mathrm{nm}$) and counterpropagating two-photon ($780+480\,\mathrm{nm}$) Doppler spreads along the optical axis evaluate to $0.039\,\mathrm{MHz}$ and $0.028\,\mathrm{MHz}$, respectively.
A frozen-position diagnostic at this thermal excursion (transverse displacement $\delta x = 266\,\mathrm{nm}$, $\theta \approx 4.5^\circ$) evaluated across five projection sectors ($M_{\mathrm{tot}}=1/2$ to $9/2$, 512 pair modes) yields a nominal fidelity of $99.903\%$, a shift of $-8.8\times 10^{-5}$ from the single-sector projection despite a transient cross-sector excursion of $1.76\times 10^{-3}$.
This multi-sector calculation finds limited leakage at the tested out-of-plane displacement. Evaluating thermal performance for an apparatus also requires sampling continuous trajectories, Doppler detunings, and the trap and beam geometry.

\subsection{Sensitivity to stray electric and magnetic bias fields}
\label{app:bias_fields}
Because the pair resonance condition and single-qubit addressing rely on Zeeman and Stark splittings, we evaluated gate sensitivity to electric- and magnetic-field offsets.
The magnetic-field scan tracks the atomic Zeeman transitions and recalibrates local-$Z$ corrections at each field [Fig.~\ref{fig:forster_robustness}(d)], showing an operating plateau near $B=3.10\,\mathrm{G}$ with infidelities below $10^{-3}$; this scan characterizes operation with recalibration, rather than noise tolerance at fixed optical frequencies.
In particular, shifting the bias by $+2.5\,\mathrm{mG}$ (from $3.1000$ to $3.1025\,\mathrm{G}$) alters the sampled joint minimum by $+5.94\times10^{-7}$ in the reference basis and $-1.13\times10^{-6}$ in the expanded $\pm60\,\mathrm{GHz}$ pair window. These shifts indicate smooth behavior near the chosen operating point.

We tested uncompensated stray electric fields with optical carriers and local-$Z$ phases held at their zero-electric-field calibration values.
Because Rydberg polarizabilities scale as $n^{*7}$, differential Stark shifts are dominated by the addressed Rydberg levels, with scalar polarizabilities governing the low-field quadratic response $\Delta E \simeq -\frac{1}{2}\alpha_0 \lvert\mathbf{E}\rvert^2$; transverse field components $E_\perp=\sqrt{E_x^2+E_y^2}$ thus induce equivalent second-order Stark shifts to axial fields $E_z$.
Across the stray-field magnitude range $\lvert\mathbf{E}\rvert\le3\,\mathrm{mV/cm}$ (tested along the quantization axis as $E_z$), the gate fidelity remains above $99.91\%$ (changing by less than $6\times10^{-7}$ in the reference basis and $3\times10^{-7}$ in the $\pm60\,\mathrm{GHz}$ test).
Even with a residual uncompensated field of $\lvert\mathbf{E}\rvert=0.01\,\mathrm{V/cm}$, the fidelity remains above $99.90\%$ ($99.9098\%$, a penalty below $3\times 10^{-6}$), before uncompensated Stark detuning degrades the resonance at larger fields ($\gtrsim 0.05\,\mathrm{V/cm}$).
The field scans indicate that compensating residual fields to the $10\,\mathrm{mV/cm}$ level preserves near-nominal fidelity in the tested model, without deliberate Stark tuning.

\subsection{Spectroscopic uncertainty and experimental calibration tolerances}
\label{app:spectroscopy_sensitivity}
Atomic level energies and quantum defects in our model are fixed by the $\mathrm{Yb171\_mqdt}$ v1.4 database~\cite{Peper2024,Kuroda2025}.
While numerical basis convergence is established to within $10^{-6}$ (Table~\ref{tab:p04_convergence}), the physical atomic levels retain empirical spectroscopic uncertainties.
As deterministic sensitivity scales representative of published MQDT fits, we adopt the $2.3\,\mathrm{MHz}$ series-level rms residual for \revsym{$6s\,nS$} Rydberg states alongside the $3.16\,\mathrm{MHz}$ deviation ($3.2\,\mathrm{MHz}$ rounded scale) reported for the tabulated entry neighboring our target $P$ state~\cite{Peper2024}.

To test sensitivity to spectroscopic uncertainties, we evaluated performance across the four signed energy-shift vertices:
\begin{equation}
(\delta E_S/h,\delta E_P/h)\in\{\pm2.3\}\times\{\pm3.2\}\,\mathrm{MHz},
\end{equation}
yielding asymptotic pair defect shifts of \revsym{$\delta\Delta_{\mathrm{F}}/h$}$=(\delta E_S-\delta E_P)/h\in\{\pm0.9,\pm5.5\}\,\mathrm{MHz}$.
In physical implementations, single-atom spectroscopy directly references optical laser carriers to the physical transitions, while in-situ calibrations optimize the virtual local-$Z$ phases.
Under carrier tracking and per-vertex local-$Z$ recalibration, the gate fidelities at the four vertices range between $99.06\%$ and $99.90\%$.
When holding local-$Z$ phases fixed at their nominal values under tracked carriers, fidelities span $97.43\%$ to $99.86\%$.

A pure target pair-defect scan holds both optical carriers and virtual-$Z$ phases fixed in the tracked atomic frame.
At detuning offsets of $\pm1\,\mathrm{MHz}$, the gate fidelities remain above $99.84\%$ ($99.85\%$ and $99.84\%$), and exceed $99.62\%$ across a wider $\pm2\,\mathrm{MHz}$ span ($99.66\%$ at $-2\,\mathrm{MHz}$ and $99.62\%$ at $+2\,\mathrm{MHz}$).
The fidelity is locally quadratic near its nominal extremum: based on the calculated local curvature, constraining residual post-calibration defect uncertainty to $\lvert\delta\Delta\rvert/h\le 0.1\,\mathrm{MHz}$ ($\le 0.2\,\mathrm{MHz}$) incurs an overlap penalty of $\approx 7\times 10^{-6}$ ($2.7\times 10^{-5}$).
In physical experiments, single-atom ultraviolet spectroscopy around $302\,\mathrm{nm}$ references optical carriers, while two-atom F\"orster spectroscopy at $R\approx 3.4\,\mu\mathrm{m}$ determines the physical defect and bright-state splitting.
For sub-megahertz deviations, adjusting the bias magnetic field near $B=3.10\,\mathrm{G}$ [with differential Zeeman sensitivity $d(\Delta_{\mathrm{F}}/h)/dB \approx 1.73\,\mathrm{MHz/G}$] allows in-situ tuning without re-optimizing the pulse. For multi-megahertz discrepancies between database predictions and experiment, the calibration would also update the central detuning command $\bar\Delta(t)$ alongside carrier tracking.
These calibration steps would need to be tested using the measured pair spectrum.

\section[A candidate static van der Waals interaction]{\texorpdfstring{\revd{A candidate static van der Waals interaction}}{A candidate static van der Waals interaction}}
\label{sec:vdw}

\begin{figure}[!tbp]
\centering
\includegraphics[width=\columnwidth]{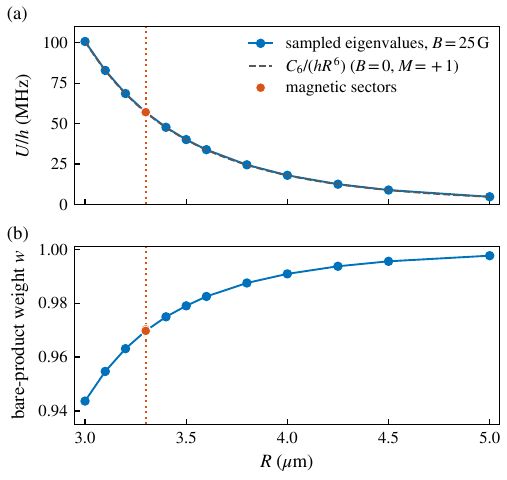}
\caption{\revd{\textbf{A sampled repulsive van der Waals branch provides a candidate static interaction scale $U/h=57.1\,\mathrm{MHz}$ at $R=3.3\,\mu\mathrm{m}$.}}
(a)~Pair shift $U/h$ for $^{87}\mathrm{Rb}\,66S_{1/2}+\mathrm{Yb}\,S(\nu{=}62.6823)$ at $B=25\,\mathrm{G}$, $\theta=0$, $(m_{\mathrm{Rb}},m_{\mathrm{Yb}})=(+1/2,+1/2)$.
Solid: maximum-overlap eigenvalues at the sampled separations.
Dashed: zero-field stretched-channel guide $C_6/(h R^6)$ with $C_6/h=+73.58\,\mathrm{GHz}{\cdot}\mu\mathrm{m}^{6}$.
The four nearly superposed markers at $R=3.3\,\mu\mathrm{m}$ denote the magnetic product configurations (spread $0.6\%$ in $U$).
The dotted vertical line marks the candidate static-interaction distance $R=3.3\,\mu\mathrm{m}$.
(b)~Bare-product pair-state weight $w$ of the tracked branch along the same finite-field track. These static diagnostics do not establish a driven gate fidelity.}
\label{fig:vdw}
\end{figure}

As a complement to the exchange-assisted F\"orster resonance, this appendix characterizes an auxiliary, selection-rule-allowed repulsive van der Waals channel between $^{87}\mathrm{Rb}(66S_{1/2})$ and $^{171}\mathrm{Yb}[S(\nu{=}62.6823)]$.
We calculate its pair shift, bare-state purity, basis convergence, and optical excitation requirements at tweezer-array spacings.
Multichannel pulse design for a blockade gate is left for future work.

The ytterbium state is the mixed-character MQDT root with effective principal quantum number $\nu\approx 62.6823$ (database root $\nu=62.682292758$, designated as $\nu=62.682293$ in Table~\ref{tab:state_dictionary}; $F=1/2$), selected in the $\mathrm{Yb171\_mqdt}$ v1.4 database as $(n{=}67, \ell{=}0, s{=}0, f{=}1/2)$~\cite{Hummel2024,Peper2024}.
Its term energy is $50415.288\,\mathrm{cm}^{-1}$ ($50415.28796\,\mathrm{cm}^{-1}$ in the database; Table~\ref{tab:state_dictionary}), corresponding to a $^{3}P_0\to S$ optical transition wavelength of $301.87\,\mathrm{nm}$ ($301.8699\,\mathrm{nm}$ in vacuum).
Following the conventions of Appendix~\ref{app:state_dictionary}, we identify this state by its effective quantum number $\nu$ and term energy.

At each sampled separation, we track the eigenstate of maximum overlap on the bare product $\lvert 66S,m_{\mathrm{Rb}}\rangle\otimes\lvert S(\nu),m_{\mathrm{Yb}}\rangle$ and define its shift relative to the separated, field-dressed atoms:
\begin{equation}
U(R)=E(R,B)-E(\infty,B),
\label{eq:Uvdw}
\end{equation}
where $E(\infty,B)$ is the energy of the separated atoms at the same applied magnetic field.
Subtracting $E(\infty,B)$ isolates the pair interaction from the atomic Zeeman shifts; a positive $U(R)>0$ corresponds to a repulsive interaction.
In the asymptotic zero-field limit ($B=0$), the nearest dipole-coupled pair channel lies $315\,\mathrm{MHz}$ below the pair asymptote.
Because the dominant coupled virtual states lie lower in energy, the second-order perturbation sum yields a positive dispersion coefficient $C_6$.
For the stretched pair states ($M=m_{\mathrm{Rb}}+m_{\mathrm{Yb}}=\pm1$), second-order perturbation theory gives $C_6/h=+73.579\,\mathrm{GHz}{\cdot}\mu\mathrm{m}^{6}$.
In the degenerate $M=0$ subspace $\{\lvert-1/2,+1/2\rangle,\lvert+1/2,-1/2\rangle\}$, the effective $C_6/h$ matrix has diagonal entries $73.607\,\mathrm{GHz}{\cdot}\mu\mathrm{m}^{6}$ and off-diagonal entries $0.056\,\mathrm{GHz}{\cdot}\mu\mathrm{m}^{6}$, yielding eigenvalues $73.551$ and $73.664\,\mathrm{GHz}{\cdot}\mu\mathrm{m}^{6}$, with little variation across the zero-field magnetic configurations.

To spectroscopically resolve a specific pair combination among the magnetic product configurations, an axial bias magnetic field of $B=25\,\mathrm{G}$ ($\theta=0$) is applied.
At this field, the calculated $\mathrm{Rb}(66S)$ $m_J$ Zeeman splitting is $70.06\,\mathrm{MHz}$ and the differential separation between the two $\mathrm{Yb}$ $\pi$ transitions is $14.04\,\mathrm{MHz}$, providing spectral separation for addressing the stretched $(m_{\mathrm{Rb}},m_{\mathrm{Yb}})=(+1/2,+1/2)$ sector.

Figure~\ref{fig:vdw} presents the finite-field pair potential obtained from full multichannel diagonalization using the reference PairInteraction setup defined in Sec.~\ref{sec:forster_channel}, with cutoffs $\Delta n=3$, $\ell\le 3$, atomic and pair energy windows of $\pm80\,\mathrm{GHz}$, and dipole--dipole coupling.
At $\theta=0$, the total azimuthal angular momentum projection $M$ is strictly conserved; the addressed $M=+1$ block contains $16\,279$ states, while the two $M=0$ product configurations belong to the same symmetry sector, with basis sizes of approximately $17\,600$ states in this electronic-basis evaluation.
At each interatomic separation $R$, we track the eigenstate of maximum overlap on the bare product $\lvert 66S,m_{\mathrm{Rb}}\rangle\otimes\lvert S(\nu),m_{\mathrm{Yb}}\rangle$, with its bare-product weight $w$ plotted in Fig.~\ref{fig:vdw}(b).
Across the 12 sampled separations between $R=3.0$ and $5.0\,\mu\mathrm{m}$, the maximum-overlap eigenvalue tracks the zero-field $C_6/(h R^6)$ curve to within $0.11\,\mathrm{MHz}$.
At a representative tweezer spacing of $R=3.3\,\mu\mathrm{m}$, the branch provides a repulsive interaction shift of $U/h=57.1\,\mathrm{MHz}$ ($57.072\,\mathrm{MHz}$) with a bare-product weight $w=0.970$. The four magnetic product configurations have an energy-shift spread of $0.6\%$ ($0.34\,\mathrm{MHz}$).

\begin{reveblock}
\begin{table}[t]
\caption{\textbf{One-at-a-time basis checks for the static vdW candidate.}
The reference uses Rb $n=66\pm3$, Yb $\nu=62.6823\pm3.3$, $\ell_{\max}=3$, and a $\pm80\,\mathrm{GHz}$ pair window.
Each other row expands one axis while keeping $B=25\,\mathrm{G}$, $R=3.3\,\mu\mathrm{m}$, $\theta=0$, total $M=+1$, and dipole--dipole coupling fixed.
$N$ is the full finite-field pair-basis size and $w$ the bare-pair weight.}
\label{tab:vdw_convergence}
\scriptsize
\begin{ruledtabular}
\begin{tabular}{lrrrr}
Setting & $N$ & $C_6/h$ & $U/h$ & $w$ \\
 & & $(\mathrm{GHz}\cdot\mu\mathrm{m}^6)$ & $(\mathrm{MHz})$ & \\
\hline
Reference & 16\,279 & 73.57945 & 57.07205 & 0.969718 \\
Rb $n=66\pm4$ & 19\,317 & 73.57948 & 57.07751 & 0.969709 \\
Yb $\nu=62.6823\pm4.3$ & 19\,522 & 73.57873 & 57.07173 & 0.969719 \\
$\ell_{\max}=4$ & 33\,180 & 73.57945 & 57.06934 & 0.969732 \\
Pair window $\pm120\,\mathrm{GHz}$ & 21\,681 & 73.58994 & 57.07991 & 0.969718 \\
\end{tabular}
\end{ruledtabular}
\end{table}

To test sensitivity to basis truncation, Table~\ref{tab:vdw_convergence} reports one-at-a-time expansions of the basis cutoffs at $R=3.3\,\mu\mathrm{m}$ and $B=25\,\mathrm{G}$.
Expanding the Rb radial cutoff to $n=66\pm4$ ($N=19\,317$), the Yb effective range to $\nu=62.6823\pm4.3$ ($N=19\,522$), the maximum orbital angular momentum to $\ell_{\max}=4$ ($N=33\,180$), or the pair-energy window to $\pm120\,\mathrm{GHz}$ ($N=21\,681$) gives maximum shifts from the reference of $0.0105\,\mathrm{GHz}\cdot\mu\mathrm{m}^{6}$ in $C_6/h$, $0.0079\,\mathrm{MHz}$ in $U/h$, and $1.31\times10^{-5}$ in $w$.
These correspond to relative variations of $1.43\times10^{-4}$, $1.38\times10^{-4}$, and $1.35\times10^{-5}$, respectively.
Contracted-basis checks and full-precision records are archived in the repository manifests. These checks assess convergence within the dipole--dipole model.
\end{reveblock}

Optical excitation of the target pair state can be implemented using standard laser wavelengths.
The rubidium transition from $\lvert 5S_{1/2}, F{=}2, m_F{=}+2\rangle$ to $\lvert 66S_{1/2}, m_J{=}+1/2\rangle$ is driven via a two-photon $780{+}480\,\mathrm{nm}$ pathway.
The ytterbium state is driven from $\lvert {}^{3}P_0, F{=}1/2, m_F{=}+1/2\rangle$ to $\lvert S(\nu{=}62.6823), F{=}1/2, m_F{=}+1/2\rangle$ using a $\pi$-polarized $301.87\,\mathrm{nm}$ beam.
With the calculated optical dipole matrix element of $\lvert d_0\rvert=6.27\times10^{-4}\,ea_0$, achieving a target Rabi frequency of $\Omega/2\pi=2\,\mathrm{MHz}$ requires approximately $18.7\,\mathrm{mW}$ in a $12\,\mu\mathrm{m}$ beam waist.
Under ideal $\pi$ polarization, a field-dressed single-atom calculation including all $S$ and $D$ lines within $\pm100\,\mathrm{GHz}$ places the nearest spectator transition more than $2.36\,\mathrm{GHz}$ from the target line.
For an illustrative resonant $2\,\mathrm{MHz}$ square $2\pi$ pulse, the calculated off-resonant excitation of spectator Rydberg states is $1.1\times10^{-5}$, while off-resonant excitation from the opposite $^{3}P_0$ Zeeman component is $1.52\times10^{-3}$, imparting a deterministic phase of $-0.20\,\mathrm{rad}$ that can be included in composite pulse design.
For the $\mathrm{Rb}(66S_{1/2})$--$\mathrm{Yb}[S(\nu{=}62.6823)]$ channel, these calculations establish a well-isolated, repulsive interaction ($U/h=57.1\,\mathrm{MHz}$ at $3.3\,\mu\mathrm{m}$) with clean spectral selectivity, providing a viable static interaction resource for hybrid architectures.

\begin{acknowledgments}
    This work was partially supported by the National Key R\&D Program of China (Grant No.~2024YFB4504004), the National Natural Science Foundation of China (Grant Nos.~12404568 and 92576114), the Quantum Science and Technology--National Science and Technology Major Project (Grant No.~2021ZD0301703), the Guangdong Provincial Quantum Science Strategic Initiative (Grant Nos.~GDZX2403008 and GDZX2503001), the Shenzhen Science and Technology Program (Grant No.~KQTD20200820113010023), and the Guangdong Provincial Key Lab of Integrated Communication, Sensing and Computation for Ubiquitous Internet of Things (Grant No.~2023B1212010007).
\end{acknowledgments}

\section*{Data and code availability}
All numerical datasets, simulation and reproduction scripts, and database provenance manifests supporting the findings of this study are openly available in Zenodo at \href{https://doi.org/10.5281/zenodo.22691580}{doi:10.5281/zenodo.22691580}~\cite{ZenodoArchive} and in the GitHub repository at \url{https://github.com/wanda0929/RbYb_pub}.
The repository contains the serialized outputs for the Förster channel characterization, driven gate trajectories under radiative decay, parametric robustness sweeps, and the static van der Waals pair track, with the figure- and table-to-data mapping documented in the repository metadata.
Stand-alone Python scripts reproduce the pair-state diagonalizations, pulse propagation, and numerical convergence audits using version-locked dependencies and archived PairInteraction atomic database manifests.
\bibliography{references}

\end{document}